\documentclass[prd,twocolumn,amsmath,amssymb,floatfix,superscriptaddress]{revtex4-1}

\usepackage{graphicx}
\usepackage{amssymb}
\usepackage{amsmath}
\usepackage{hyperref}

\usepackage{color}

\DeclareMathAlphabet\mathbfcal{OMS}{cmsy}{b}{n}
\definecolor{darkgreen}{cmyk}{0.85,0.2,1.00,0.2} 
\definecolor{purple}{cmyk}{0.5,1.0,0,0}

\def\barray{\begin{array}} 
\def\earray{\end{array}}
\def\be{\begin{equation}}
\def\ee{\end{equation}}
\def\ben{\begin{equation} \nonumber}
\def\een{\end{equation}}
\def\ban{\begin{eqnarray*}}
\def\ean{\end{eqnarray*}}
\def\ba{\begin{eqnarray}}
\def\ea{\end{eqnarray}}

\def\({\left(}
\def\){\right)}

\newcommand{\simgt}{\lower.5ex\hbox{$\; \buildrel > \over \sim \;$}}
\newcommand{\simlt}{\lower.5ex\hbox{$\; \buildrel < \over \sim \;$}}

\newcommand{\mnras}{Mon.~Not.~Roy.~Astron.~Soc.}

\newcommand{\pasj}{PASJ}

\newcommand{\bq}{{\bf q}}

\newcommand{\Omegam}{{\Omega_\textrm{m}}}
\newcommand{\Omegab}{{\Omega_\textrm{b}}}
\newcommand{\Omegac}{{\Omega_\textrm{c}}}
\newcommand{\OmegaL}{{\Omega_\Lambda}}

\newcommand{\As}{{A_\textrm{s}}}
\newcommand{\lnAs}{{\ln\!A_\textrm{s}}}
\newcommand{\ns}{{n_\textrm{s}}}
\newcommand{\br}{{\rm b}}
\newcommand{\hMpci}{$h\,$Mpc$^{-1}$}
\newcommand{\hiMpc}{$h^{-1}\,$Mpc}

\newcommand{\Dv}{{\cal P}}

\begin{document}

\title{Super-Sample Signal}
\author{Yin Li}
\affiliation{Department of Physics, University of Chicago, Chicago, Illinois 60637, U.S.A.}
\affiliation{Kavli Institute for Cosmological Physics, Department of Astronomy \& Astrophysics,  Enrico Fermi Institute, University of Chicago, Chicago, Illinois 60637, U.S.A.}
\author{Wayne Hu}
\affiliation{Kavli Institute for Cosmological Physics, Department of Astronomy \& Astrophysics,  Enrico Fermi Institute, University of Chicago, Chicago, Illinois 60637, U.S.A.}
\author{Masahiro Takada}
\affiliation{Kavli Institute for the Physics and Mathematics of the Universe
(WPI),
Todai Institutes for Advanced Study,
The University of Tokyo, Chiba 277-8583, Japan}

\begin{abstract}
When extracting cosmological information from power spectrum measurements,
we must consider the impact of super-sample density fluctuations whose wavelengths are larger than the survey scale.  
These modes contribute to the mean density fluctuation $\delta_\br$  in the survey and change the power spectrum in the same way as a change in the cosmological background.  They can  be simply included
in cosmological parameter estimation 
and 
forecasts 
by treating $\delta_\br$ 
as an additional cosmological parameter enabling efficient exploration of its impact.
To test this approach,  we consider here an idealized measurement of the matter power spectrum itself in
the $\Lambda$CDM cosmology though our 
 techniques can readily be extended to more observationally relevant statistics or
 other parameter spaces.  Using sub-volumes of large-volume
$N$-body simulations for power spectra measured with respect to either the global or local mean density, we  verify that the minimum  variance estimator of $\delta_\br$ is both unbiased and has the predicted
variance.
Parameter degeneracies arise since the response of the matter power spectrum to $\delta_\br$
and cosmological parameters share similar properties in changing the growth of structure and
dilating the scale of features  especially in the local case.   For matter power spectrum measurements, these degeneracies can lead in certain cases to substantial error degradation 
and motivates future studies of specific cosmological observables such as galaxy clustering
and weak lensing statistics with these techniques.
\end{abstract}

\maketitle

\section{Introduction}

The statistical properties of large scale structure provide a
wealth of cosmological information on fundamental physics, including
cosmic acceleration, neutrino masses and inflation. The simplest
statistic is the two-point correlation function or power spectrum of the
matter density field which underlies observables such as weak lensing
and galaxy clustering. To extract cosmological information from  the power spectrum of upcoming wide-area galaxy surveys (e.g.~\cite{TakadaPFS:14}), we need to properly model its own statistical 
properties, one of which is the impact of modes whose wavelengths are
larger than the survey scale, the so-called super-sample modes.

  While these super-sample modes are not
directly observable, they impact the evolution of sub-sample modes in an
observable way due to nonlinear mode coupling
\citep{Hamiltonetal:06,Sefusattietal:06,HuKravtsov:03,TakadaBridle:07,TakadaJain:09,Satoetal:09,Takahashietal:09,dePutter:2011ah,Schneider:2011wf,Kayoetal:13,TakadaSpergel:13,Takada:2013wfa,Lietal:14,Schaanetal:14}.
Ref.~\cite{Takada:2013wfa} developed a simple, unified approach that describes
 the impact of the super-sample modes as the response of the power spectrum to a change in
the mean density in the finite-volume region.  Ref.~\cite{Lietal:14} then utilized
the so-called {separate universe approach} to calibrate this response in $N$-body simulations.   Here,  the mean density fluctuation $\delta_\br$ is
absorbed into   a change in cosmological
parameters
{(e.g.~\cite{TormenBertschinger:96,Cole:97,Sirko:05,Baldauf:2011bh,Gnedin:2011kj}).}    The separate universe approach
 is also useful for gaining a physical understanding of the response.  In a coherently overdense region structure grows more quickly and regions expand less quickly.   The response is
 therefore a change in the amplitude or growth of structure and in the scale or dilation
 of features in the power spectrum.
 
 The impact of the mean density fluctuation can also be viewed in two ways: as additional
 noise due to its stochasticity, or as additional signal from which the mean density in
 a given volume may be recovered \cite{Takada:2013wfa}.     In the former view, 
 the super-sample effect introduces a covariance to power spectrum estimates since
 each realization of  $\delta_\br$ coherently changes the power spectrum
 according to the calibrated response.   Ref.~\cite{Lietal:14} used the covariance
 matrix of power spectra in  subsampled simulations to verify the separate universe
 response itself.    
 
 Here we develop the alternative view that in each realization the
 super-sample effect biases the measured power spectrum or equivalently introduces
 an extra parameter, $\delta_\br$, upon which it depends.
 The mean density fluctuation leaves a signal in the power spectrum
 which can be used to recover its value providing constraints on super-sample modes
 that are not directly observable in the survey {(see also
 \cite{TakadaSpergel:13,Chiangetal:14}).}  
Moreover, this view facilitates studies
 of the impact of super-sample modes on cosmological parameter estimation by treating
 the two on an equal footing.

While here we only test this view in the idealization that the matter power spectrum is
directly observed, these ideas can be readily extended to more observationally relevant
power spectrum observables once their response to $\delta_\br$ is calibrated.

The structure of this paper is as follows. In \S~\ref{sec:sss} we
 review the super-sample effects on the power spectrum and test its interpretation as signal
 by explicitly constructing an estimator of the mean density fluctuation in simulations.
In \S~\ref{sec:degen} we study the  similarities between the power spectrum response to the mean density fluctuation and to cosmological parameters which propagate into parameter
degeneracies in the Fisher information matrix.   We discuss these results in \S~\ref{sec:discussion}.   Appendix~\ref{app:sim} provides details of the simulations and the numerical techniques.

\section{Super-sample signal}
\label{sec:sss}

In  \S~\ref{sub:signal}, we first briefly review the impact of super-sample modes on the power spectrum covariance
 which was developed in 
{Refs.~\cite{Takada:2013wfa,Lietal:14}}.   Next we
 show the effect can alternately be treated as a signal that allows us to construct an unbiased
 minimum variance estimator of the mean density fluctuation 
in  \S~\ref{sub:est1}.
We test this estimator against cosmological simulations in \S~\ref{sub:sigtest}
and show that its statistical properties are well-characterized by the separate universe response of the
power spectrum to the mean density and the covariance of the power spectrum in the
absence of the mean density mode.

\begin{figure}[tb]
    \centering
    \includegraphics[width=3.4in]{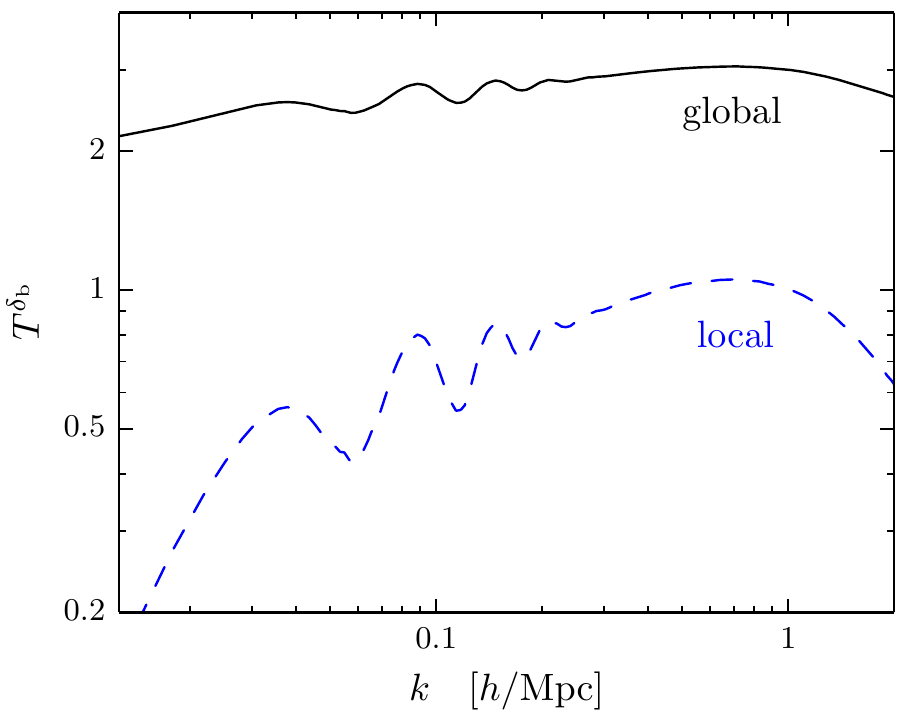}
    \caption{
Power spectrum response to the mean density fluctuation in a survey region,
calibrated with the separate universe simulations through
a change in background parameters (see Ref.~\cite{Lietal:14}).  
Solid curve represents the response of the power spectrum
referenced to the true or global mean density,
while the dashed curve to the mean survey or local density.  
The two are related by an additive constant 2,
which changes the shape of the response.
    }
    \label{fig:dPddT}
\end{figure}

\subsection{Super-sample modes}
\label{sub:signal}

In this section we characterize the effect of super-sample modes  on power spectrum measurements
in a finite survey volume in a manner that brings out their role as both signal
and noise.  Let us denote the average density fluctuation in the survey region from
these modes as $\delta_\br$ and assume that the measurement of the power spectrum of sub-survey modes  $\hat P(k;\delta_\br)$ is sample variance limited.   Let us bin these measurements into
a data vector 
\begin{equation}
\hat \Dv_i(\delta_\br) \equiv \frac{k_i^3 \hat P(k_i; \delta_\br)}{2\pi^2},
\label{eq:Di_def}
\end{equation}
and consider the impact of $\delta_\br$ on the mean and covariance of these data.
We use the dimensionless power spectrum since it is independent of the units
with which the wavenumber is measured and thus allows a cleaner separation
of effects on the amplitude and scale of structure (see Ref.~\cite{Lietal:14}).

Even if we define an estimator which yields the true power spectrum $P(k_i)$
when averaged over realizations of all modes, including the super-sample ones
\begin{equation}
\langle \hat \Dv_i \rangle \equiv \Dv_i = \frac{k_i^3  P(k_i)}{2\pi^2},
\label{eq:ave_Di}
\end{equation}
the average over realizations of different sub-survey modes 
at {\em fixed} $\delta_\br$, which we denote by
$\langle \ldots \rangle_\br$, is biased
\begin{equation}
    \langle \hat \Dv_i \rangle_\br \equiv \Dv_i(\delta_\br) = \Dv_i (1 + T^{\delta_\br}_i \delta_\br).
\label{eq:psest_sub}
\end{equation}
Here we have linearized the response of the power spectrum assuming 
$|\delta_\br|\ll 1$
\begin{equation}
T^{\delta_\br}_i \equiv \frac{\partial \ln \Dv_i}{\partial \delta_\br}.
\label{eq:global}
\end{equation}

As demonstrated in Ref.~\cite{Lietal:14} through its effects
on the power spectrum covariance below, we can use the separate
universe  approach to calibrate the power spectrum response,
$T^{\delta_\br}_i$, for a given fiducial cosmological model. 
{In this approach}
the mean density {fluctuation}
of the survey $\delta_\br$ is absorbed
into {the local background density} by redefining cosmological parameters.

{Fig.~\ref{fig:dPddT} summarizes the results
for the $\Lambda$CDM
model  (see Tab.~\ref{tab:LCDMpar}) at $z=0$ using $N$-body simulations to calibrate the response deep into the nonlinear regime.
    Note that the
technique itself is more general than this particular implementation.  Given a state-of-the-art simulation that includes baryonic effects, e.g.\ star formation and feedback {\cite{Rudd:2007zx,vanDaalenetal:11}}, the impact of super-sample modes can still be calibrated by runs where the background parameters are changed to absorb $\delta_\br$.

These results apply to power spectrum measurements where the global mean density
is known through cosmological parameters, e.g.\ in the case of weak
lensing power spectra.
{If the power spectrum is estimated with respect to the local mean
density within the survey region, which is the case for  galaxy surveys, the relevant
observable is}
\begin{equation}
\hat \Dv^W = \frac{\hat \Dv}{(1+\delta_\br)^2},
\label{eq:Plocal}
\end{equation}
{where $W$ denotes the survey volume or window here and below.}
The power spectrum response is therefore modified to be
\begin{equation}
T^{\delta_\br}_i  \big|_{\rm local} \equiv \frac{\partial \ln \Dv^W_i}{\partial\delta_\br} \approx \frac{\partial\ln \Dv_i}{\partial\delta_\br}-2.
\label{eq:Tlocal}
\end{equation}
Because the formalism for super-sample effects is otherwise identical, we use the
term ``local" data or response when replacing the 
``global" versions of Eqs.~(\ref{eq:global}) and (\ref{eq:Plocal}), in relevant formula below.
{Both responses are shown in Fig.~\ref{fig:dPddT}.}

Since $\delta_\br$ is 
a random zero mean variable, its variance turns into a 
 covariance of the power spectrum estimators. 
  In this
 sense the super-sample effect contributes as additional noise.  We can model this noise
 by treating the bias as a purely
systematic additive shift in the power spectrum per realization
\begin{equation}
\hat \Dv_i(\delta_\br) = \hat \Dv_i(0) + \Dv_i  T^{\delta_\br}_i 
\delta_\br.
\label{eq:model}
\end{equation}
The covariance {matrix} of the power spectrum data
is then
given by 
\begin{equation}
\langle \hat \Dv_i \hat \Dv_j \rangle -\Dv_i \Dv_j  
= C_{ij} +\Dv_i  \Dv_j T^{\delta_\br}_i  T^{\delta_\br}_j  \sigma_\br^2,
\label{eq:cov1}
\end{equation}
where $\sigma^2_\br$ is the variance of the mean density field in the
survey window, defined as
\begin{equation}
\sigma_\br^2 \equiv  \langle \delta_\br^2 \rangle
=\frac{1}{V_W^2}\int\!\!\frac{d^3\bq}{(2\pi)^3}|\tilde{W}(\bq)|^2
{P(q).}
\label{eq:sigma}
\end{equation}
{Here $V_W$ is the survey volume and
$\tilde{W}(\bq)$ is the survey window function which acts as a low pass
filter.
{For 
 a sufficiently large survey volume, the super-sample modes are in the linear regime
 and therefore
$\sigma_\br$ can be accurately calculated with the linear power spectrum either by explicit computation of Eq.~(\ref{eq:sigma}) or Gaussian realizations of the linear
density field.} 
The matrix $C_{ij}$ in Eq.~(\ref{eq:cov1}) is defined as}
\begin{equation}
C_{ij} =  \langle \hat \Dv_i(0) \hat \Dv_j(0) \rangle - \Dv_i \Dv_j. 
\end{equation}
Note that this is the covariance of the power spectrum estimators in the absence of
the super-sample effect.
It  can be readily
calibrated from a suite of small-volume
$N$-body simulations 
for {a}
given cosmological
model \cite{Scoccimarroetal:99,MeiksinWhite:99} (see Appendix \ref{app:sim} and Ref.~\cite{Lietal:14} for implementation specifics). The sum of the two contributions of Eq.~(\ref{eq:cov1})
reproduces the super-sample covariance derived in Ref.~\cite{Lietal:14} from 
trispectrum considerations. 

An alternate model would be to consider the bias {in the power
spectrum estimators}
to be multiplicative with respect to $\hat  \Dv_i(0)$,
\begin{equation}
\hat \Dv_i = \hat \Dv_i(0) (1 +  T^{\delta_\br}_i \delta_\br).
\label{eq:multiplicative}
\end{equation}
{This model yields an additional contribution to the second term of Eq.~(\ref{eq:cov1})
\begin{equation}
\Dv_i  \Dv_j T^{\delta_\br}_i  T^{\delta_\br}_j  \sigma_\br^2 \rightarrow (\Dv_i  \Dv_j + C_{ij}) 
T^{\delta_\br}_i  T^{\delta_\br}_j  \sigma_\br^2 
\end{equation}
and hence only a small change for well measured bins where the covariance
is much smaller than the product of the means.  We hereafter adopt the additive model.}

\subsection{Signal vs.\ noise}
\label{sub:est1}

In a given realization of the survey volume, the super-sample effect systematically
changes the power spectrum of sub-survey modes just like a cosmological parameter 
$\mathbf{p}_\textrm{c}$.
Indeed, in the additive model of Eq.~(\ref{eq:model}), the analogy is precise: 
$\delta_\br$ is simply a parameter
that changes the mean power spectrum
\begin{equation}
 \hat \Dv_i(\delta_\br) -
 \Dv_i(\delta_\br ,\mathbf{p}_\textrm{c}) 
 =  \hat \Dv_i(0)  - \Dv_i(\mathbf{p}_\textrm{c}) .
\end{equation}
Thus the data in the presence of $\delta_{\br}$ has the same statistical properties 
as in its absence.  They are both drawn from a distribution with covariance $C_{ij}$ and only
the mean is shifted.
Parameter estimation then proceeds in the usual way by treating $\delta_{\br}$ as
a parameter $\mathbf{p}=\{\delta_\br,\mathbf{p}_\textrm{c}\}$.
For example, the posterior
probability of model parameters including $\delta_\br$  can be estimated using
 Markov Chain Monte Carlo techniques based on the likelihood function constructed from the covariance matrix $C_{ij}$ with the model parameterized
by $\mathbf{p}$.
In this view, the super-sample effect is a signal which allows the mean density fluctuation
to be recovered rather than a source of additional noise.

To test this interpretation with simulations, we can construct an explicit estimator of $\delta_\br$ based on  linear response.
Here
we shall assume that cosmological parameters are fixed and hence suppress their
appearance in the expressions.
A general  unbiased linear estimator {of $\delta_\br$} takes the form
\begin{equation}
    \hat\delta_\br = \sum_i w_i (\hat \Dv_i - \Dv_i),
    \label{eq:est1}
\end{equation}
where the weight 
\begin{equation}
\sum_i w_i \Dv_i T_i^{\delta_\br} = 1,
\end{equation}
is constrained by the condition 
$\langle {\hat\delta_\br} \rangle_\br = \delta_\br$.   The remaining freedom in choosing the
weights is fixed by minimizing the variance of the estimator 
\begin{equation}
   {\sigma_{\delta_\br}^2 \equiv}
{\langle \hat \delta_\br^2 \rangle}_\br -\delta_\br^2
    =  \sum_{ij} w_i w_j C_{ij},
    \label{eq:estvar}
\end{equation}
subject to the Lagrange multiplier constraint yielding
\begin{equation}
    w_i = \frac{\sum_{j} T_j^{\delta_\br} \Dv_j \left[{\bf C}^{-1}\right]_{ij}}
        {\sum_{jk} T_j^{\delta_\br}\Dv_j \left[{\bf C}^{-1}\right]_{jk} T_k^{\delta_\br} \Dv_k}
    \label{eq:w1}
\end{equation}
and
\begin{equation}
\sigma_{\delta_\br}^2
= \left( \sum_{jk} T_j^{\delta_\br}\Dv_j \left[{\bf C}^{-1}\right]_{jk} T_k^{\delta_\br} \Dv_k \right)^{-1}.
\label{eq:bvar}
\end{equation}
If instead of the additive bias model, the multiplicative bias model of  Eq.~(\ref{eq:multiplicative}) is correct, this estimator remains unbiased but its variance changes.  
Thus comparing the predicted variance from Eq.~(\ref{eq:estvar}) with the variance
obtained from simulations tests the accuracy of our additive model of
super-sample effects as well as 
that of the response $T^{\delta_\br}_i$ and covariance $C_{ij}$ calibration.

\begin{figure}[tb]
\centering
    \includegraphics[width=3.4in]{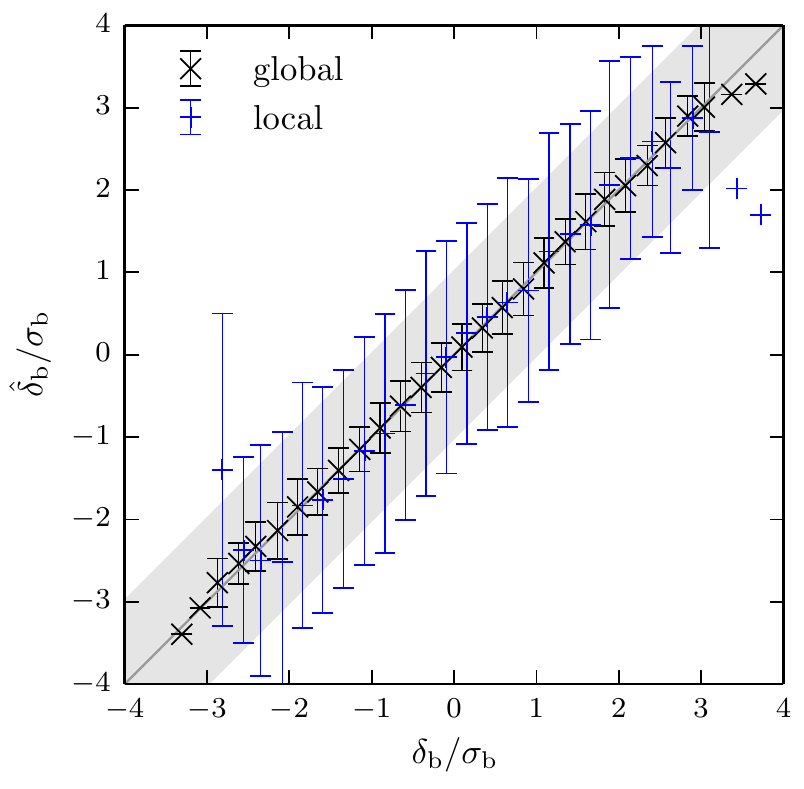}
    \caption{
%        \footnotesize 
Estimated ($\hat\delta_\br$) vs.\ true ($\delta_\br$) mean density 
{fluctuations}
    for the set of $3584$ 
simulation subvolumes.  
        The means ($\times$, $+$, slightly shifted for clarity) of the estimators using both globally and locally referenced power spectra
        show no trace of {a}
bias 
        for bins with sufficient statistics that the standard deviation
        (errorbars)  can be estimated.  
This standard deviation is
        significantly smaller than  $\pm\sigma_\br$ (gray band)  for the global case
        and comparable for the local case.  
    }
    \label{fig:sss}
\end{figure}

\subsection{Density estimation}
\label{sub:sigtest}

In this section, we test the $\hat \delta_\br$ estimator with large-volume simulations where the true 
mean density fluctuation $\delta_\br$ is known in each subvolume of the
simulation.
We summarize here the relevant simulation details presented in 
Appendix~{\ref{app:sim}}.
Specifically 7 large-volume simulations, each with a $4\,\textrm{Gpc}/h$ box length
are divided into a total of $N_\textrm{s}=7\times8^3=3584$ subvolumes of
size $500\,$\hiMpc\ each.  For reference the variance of the mean density fluctuation of the
subvolumes is
\begin{align}
\hat \sigma^2_{\br} = &   \frac{1}{N_\textrm{s}} \sum_{a=1}^{N_\textrm{s}} \delta_{\br,a}^2 = (0.01263)^2
\end{align}
 which matches well the computation from 
Eq.~(\ref{eq:sigma}), {$\sigma_\br=0.01258$}.
We measure the power spectrum {$\hat \Dv_i$}
in each subvolume separately.
Note that this power spectrum is  of the density fluctuation from the large-box mean and
hence  characterizes the ``global" data in the language of \S \ref{sub:signal}.   For the ``local"
data, we use the true average density in the subvolume
 to rescale the global data according to Eq.~(\ref{eq:Plocal}).

To calibrate the mean power spectrum $\Dv_i$ and the covariance matrix $C_{ij}$,
we run another suite of $N_\textrm{s}$ small box simulations each of the same size as the subvolume.
All power spectra are binned in $80$ $k$-bins per decade and we utilize
measurements 
out to $k\lesssim2\,$\hMpci\ {up to which we verified 
the response calibration is
accurate to several percent or better, based on
 higher-resolution simulations
\citep{Lietal:14}.}

\begin{figure}[tb]
\centering
    \includegraphics[width=3.4in]{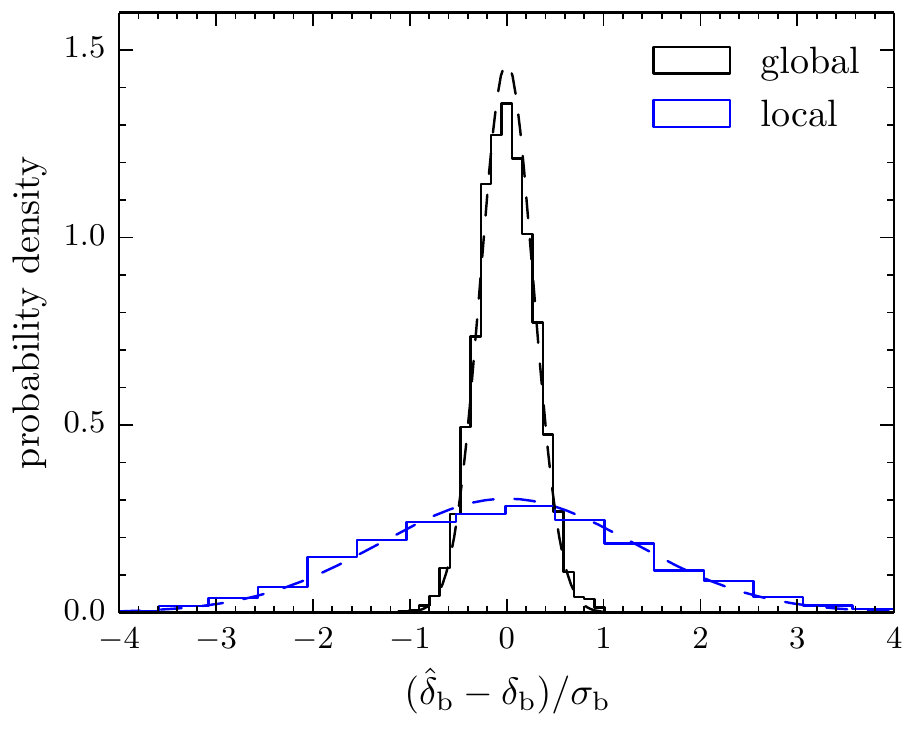}
    \caption{
%        \footnotesize 
Distribution or scatter of the density estimator $\hat\delta_\br$ around the true value  (histogram) versus the prediction from  Eq.~(\ref{eq:estvar}) with a Gaussian distribution
        (curve).   In both the global and local cases, the predictions
 are within $7\%$ of the simulation results in 
{the}
standard deviation justifying their use in parameter forecasts below.
    }
    \label{fig:estres}
\end{figure}

From  $\hat\Dv_i$, $\Dv_i$, and $C_{ij}$ we form the $\hat\delta_\br$ estimator 
of Eq.~(\ref{eq:est1}) for both the global and local cases and compare it with the 
 true value $\delta_\br$ in each of $N_\textrm{s}$ subvolumes. 
Fig.~\ref{fig:sss} shows the results binned in $\delta_\br$ with width $\sigma_\br/4$.
In both the global and local estimations, there is no trace of a bias in the estimator
to a small fraction of its standard deviation.  Moreover, in Fig.~\ref{fig:estres} we test 
the predicted distribution of the estimator combining all bins with the prediction from
Eq.~(\ref{eq:estvar}) for the variance under the assumption of a Gaussian distribution.   
The agreement in both cases is very good.   For a more quantitative assessment
we measure the variance of the estimator with our $N_\textrm{s}$ samples in the usual
way
\begin{align}
\hat \sigma^2_{\delta_\br} = &   \frac{1}{N_\textrm{s}} \sum_{a=1}^{N_\textrm{s}} (\hat\delta_{\br,a} -\delta_{\br,a})^2.
\end{align}
The result is $\hat\sigma_{\delta_\br}=0.29\sigma_\br$ for the global case 
and $1.4\sigma_\br$ for the local case.  In both cases the result is only $\sim 7\%$
larger in standard deviation than 
the prediction of Eq.~(\ref{eq:estvar}).  Even in the local case
where the standard deviation is comparable to $\sigma_\br$ there is non-negligible
extra information provided by the estimator.

These results validate the use of our additive model to predict the impact of
the super-sample effect in other cases of interest.
In Fig.~\ref{fig:errdbUmgnl}, we use this model to explore the dependence of the standard deviation of the
estimator on the maximum $k$ used.  Note that in the local case, $\sigma_{\delta_\br}$ is only a weak function of $k_{\rm max}$.   This is because, the intrinsic covariance
between bins $C_{ij}$ induces very similar changes to the power spectrum as 
$T_{\delta_\br}$ local for $k\gtrsim 0.1\,$\hMpci.    The difference in shape between
the local and global responses causes an improvement
in the standard deviation of the latter for $k\gtrsim 1\,$\hMpci.

The above results apply to estimates of $\delta_\br$ when all other parameters
that change the power spectrum are known a priori.   With joint estimator of parameters
from the survey, the $\delta_\br$ mode will degrade results on cosmological parameters
and vice versa if their impact on the power spectrum is sufficiently similar to 
cause degeneracies.  

Finally in the additive model, the scaling of these results with the volume of the survey 
is also simple.   Since $C_{ij}$ characterizes the covariance of subsurvey modes
in the absence of the super-sample effect, it scales with volume as
\begin{equation}
C_{ij}(V) = \frac{V_0}{V} C_{ij} (V_0)
\label{eq:CijV}
\end{equation}
where $V_0 = (500~h^{-1}\, {\rm Mpc})^3$, the volume of the simulation test.  Hence
\begin{equation}
\sigma_{p_{\mu}}(V) =\sqrt{ \frac{V_0}{V} } \sigma_{p_\mu} (V_0)
\label{eq:sigmaV}
\end{equation}
for any parameter estimated from power spectra data including $\delta_\br$.
The ratio $\sigma_{\delta_\br}/\sigma_\br$ does in general depend on volume.   However
in $\Lambda$CDM the quantity $\sigma_\br V^{1/2}$ varies weakly with $V$, typically by
{$\sim \sqrt{2}$ across a $V/V_0=100$.
  Thus we expect that the
relative impact of the super-sample  effect will be only weakly
dependent on volume for the cubic geometry we consider.   
{In Ref.~\cite{Takada:2013wfa} it was also shown that for a cylindrical geometry
the scaling holds although $\sigma_\br(V_0)$ itself is smaller for the same volume.}

\begin{figure}[tb]
    \centering
    \includegraphics[width=3.4in]{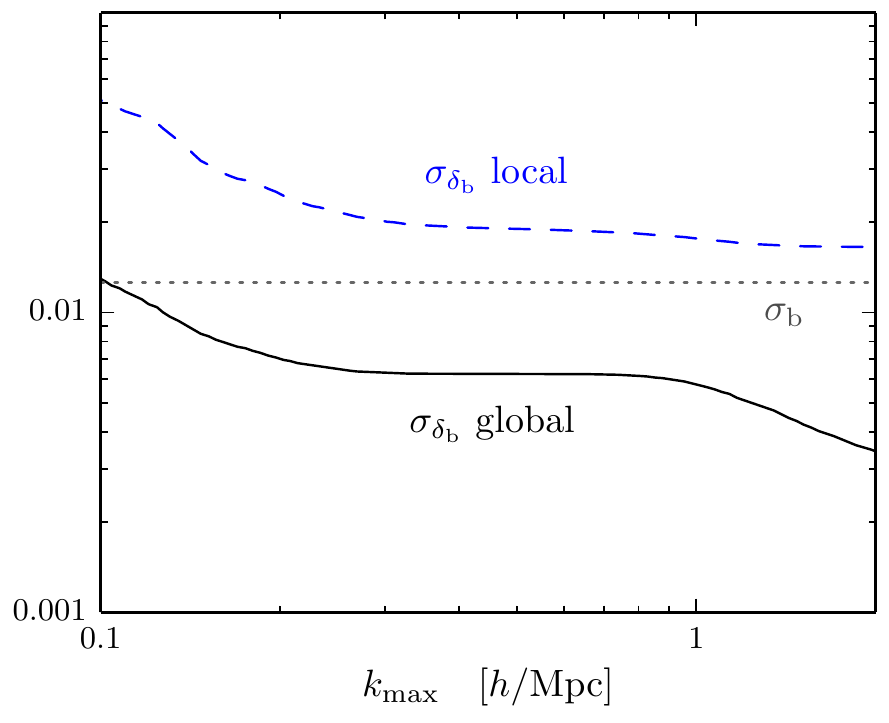}
    \caption{
        \footnotesize Standard deviation of the $\delta_\br$
 estimator, 
$\sigma_{\delta_\br}$, in the local and global cases as a function of the
        maximum $k$-bin compared with % \mtrv{the linear theory prediction} - this isn't the variance of an estimator 
the rms density fluctuation
$\sigma_\br$.  In the local
        case the standard deviation approaches the rms in the nonlinear regime and remains
        nearly constant to $2\,$\hMpci.  In the global case, it drops below the rms in the
        nonlinear regime and continues to improve beyond $1\,$\hMpci.
    }
    \label{fig:errdbUmgnl}
\end{figure}

\section{Parameter forecasts}
\label{sec:degen}

In this section we use the Fisher information matrix formalism to
study the impact of the super-sample effect on cosmological parameter
estimation in the idealized context of direct matter power spectrum measurements.  After defining the Fisher  
matrix in \S~\ref{sub:est2}, we study similarities in the power spectrum response
between the super-sample mode and cosmological parameters
in
\S~\ref{sub:Ts}.   These similarities lead to degeneracies 
which degrade errors when parameters are jointly estimated
in \S~\ref{sub:degen} and are themselves limited by 
prior information on the variance of the super-sample mode
\ref{sub:degen_prior}.

\subsection{Fisher matrix}
\label{sub:est2}

As discussed in \S \ref{sub:est1}, the additive
model $\delta_\br$ can be thought of as an additional parameter of the model
power spectrum so that the full parameter vector is 
$ \mathbf{p}\equiv\{\delta_\br,  \mathbf{p}_\textrm{c}\}$, where $\mathbf{p}_\textrm{c}$ are cosmological parameters.   In the Fisher approximation, the information
from the power spectrum mean is added to any prior information  $F_{\mu\nu}^{\rm prior}$
on the parameters,
\begin{equation}
F_{\mu\nu} = \sum_{ij} \Dv_i T_i^{\mu} [{\bf C}^{-1}]_{ij} \Dv_j T_j^{\nu} + F_{\mu\nu}^{\rm prior},
\label{eq:covp}
\end{equation}
where 
\begin{equation}
    T_i^{\mu} \equiv \frac{\partial\ln\Dv_i}{\partial p_\mu}.
\end{equation}
The inverse Fisher matrix
is an approximation to the covariance matrix of the parameters
\begin{equation}
 [{\bf F}^{-1}]_{\mu\nu}\approx \langle \hat{ p}_\mu \hat{ p}_\nu \rangle_\br
    -  p_\mu  p_\nu .
    \label{eq:paramcov}
\end{equation}
To make contact with \S \ref{sub:sigtest}, note that in the limit that the parameter space
includes only $\delta_\br$ and there is no prior information on it, Eq.~(\ref{eq:paramcov}) yields the variance
of the estimator given in Eq.~(\ref{eq:bvar}).  This is because  the Fisher approximation involves
the same linearization of the response to the parameters that permits the construction of
a linear  minimum variance unbiased estimator.
With additional parameters and no external prior, degeneracies  where the errors strongly covary appear if
the responses take a similar form.   Thus to understand the impact of super-sample modes on
parameter estimation we must compare the various responses $T_i^\mu$.

\begin{figure}[tb]
    \centering
    \includegraphics[width=3.4in]{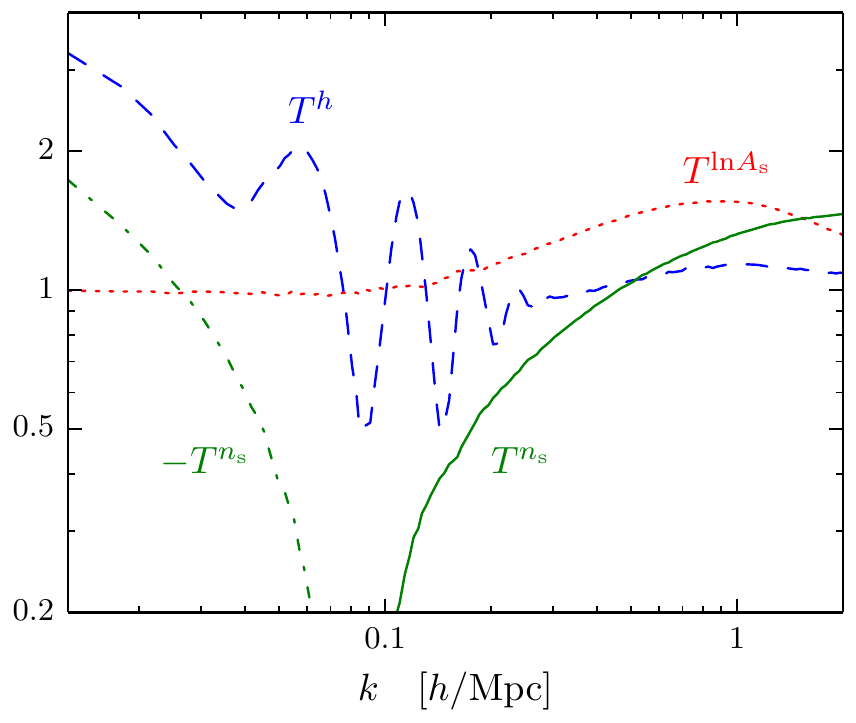}
    \caption{
        \footnotesize 
        Power spectrum response to cosmological parameters $\{ \lnAs, \ns, h \}$ 
        as calibrated from simulations.    Similarities between these responses and
        Fig.~\ref{fig:dPddT} cause degeneracies which degrade parameter errors once uncertainties
        in the super-sample mode $\delta_\br$ are marginalized.}
    \label{fig:Tp}
\end{figure}

\subsection{Parameter responses}
\label{sub:Ts}

Given that the separate universe technique described in \S \ref{sub:signal} and detailed in Ref.~\cite{Lietal:14} 
involves modeling the super-sample effect by changes in background cosmological parameters, we expect the
response to $\delta_\br$ and cosmological parameters will contain similarities that can create parameter 
degeneracies.  Namely, these parameters change the amplitude of power in the spectrum and the
scale at which  features like baryon acoustic oscillations 
(BAO)
appear.  We will call these
effects growth or ``g''  and  dilation or ``d'' respectively.

Let us begin by examining this decomposition for $\delta_\br$ and a power spectrum measurement with respect to the local mean density of the survey, defining
\begin{equation}
    T^{\delta_\br} \big|_\textrm{local} \equiv 2 \frac{\partial \ln D}{\partial \delta_\br }  T^{\delta_\br,\textrm{g}} - \frac13 T^{\delta_\br,\textrm{d}}.
    \label{eq:Tb}
\end{equation}
Here and below we omit the $k$-bin index 
$i$ where no confusion should arise.
The first term is due to the enhancement of the growth of structure in a coherently
overdense region.   Absorbing this fluctuation into a redefinition of the background implies a change in the linear growth function of
density fluctuations $D$ with respect to the {\it local} mean density 
\cite{Baldauf:2011bh}
\begin{equation}
\frac{\partial \ln D}{\partial \delta_\br } \approx \frac{13}{21}.
\end{equation}
{Hence, with this normalization,  $\lim_{k\rightarrow 0}T^{\delta_\br,
\textrm{g}}=1$.}

The second term of Eq.~(\ref{eq:Tb}) is due to the fact that an overdense region expands less quickly than the global universe.  This changes the comoving scale of physical features in the
power spectrum according to a dilation template
\begin{equation}
    T^\textrm{d} \equiv \frac{\partial \ln\Dv}{\partial \ln k}.
\label{eq:Td}
\end{equation}
The factor of $1/3$ arises since an equal time comparison is at equal 
{physical}
%local 
mean density and so
the scale factor is adjusted by $(1+\delta_\br)^{1/3}$.
By removing this rescaling with a choice of simulation box size introduced in
Ref.~\cite{Lietal:14} (see also Appendix \ref{app:sim}), we can determine $T^{\delta_\br,\textrm{g}}$ independently of
the full response $T^{\delta_\br}$ as shown in Fig.~\ref{fig:growth}.   The difference
between these responses gives the dilation template $T^{\delta_\br,\textrm{d}}$ which is
compared with the $k$-derivative of the mean power spectra $T^\textrm{d}$ in Fig.~\ref{fig:dilation}.  The agreement is good and as discussed in 
the Appendix, the response difference provides a more accurate way of calibrating
dilation than differencing noisy power spectrum data.

 Finally, when referenced to the true global mean density, the response becomes 
 (see Eq.~\ref{eq:Tlocal})
\begin{equation}
        T^{\delta_\br}  =   T^{\delta_\br} \big|_\textrm{local}  + 2.
        \label{eq:localglobal}
\end{equation}
Note that the additional factor flattens the response of the power spectrum
as shown in Fig.~\ref{fig:dPddT}.  
We shall see that this addition is important for understanding parameter degeneracies
(cf.~\cite{Mohammed:2014lja}).

Now let us compare this response with those of cosmological parameters. 
 The full
set of flat $\Lambda$CDM parameters are $\As, \ns$ which jointly define the
{primordial}
curvature power spectrum 
\begin{equation}
    {\cal P}_{\cal R} = \As \left( \frac{k}{0.05\,{\rm Mpc}^{-1}} \right)^{\ns-1};
\end{equation}
 the baryon density $\Omegab h^2$; the cold dark matter density
$\Omegac h^2$ and the dimensionless Hubble constant $h$.   The parameters $\Omegab h^2$ and 
$\Omegac h^2$ are well determined by the {cosmic microwave
background (CMB) data}
%CMB 
and additionally are not degenerate with  pure growth and dilation 
due to the changes in the BAO and matter radiation equality features that they induce.
We therefore study responses to changes in the
parameter set $\mathbf{p}_\textrm{c} = \{ \lnAs, \ns, h \}$ and keep the remaining parameters
fixed to their fiducial values.   
We calibrate each response function as discussed in Appendix~\ref{sub:calibrations} 
and show the results in Fig.~\ref{fig:Tp}. 

\begin{figure}[tb]
    \centering
    \includegraphics[width=3.4in]{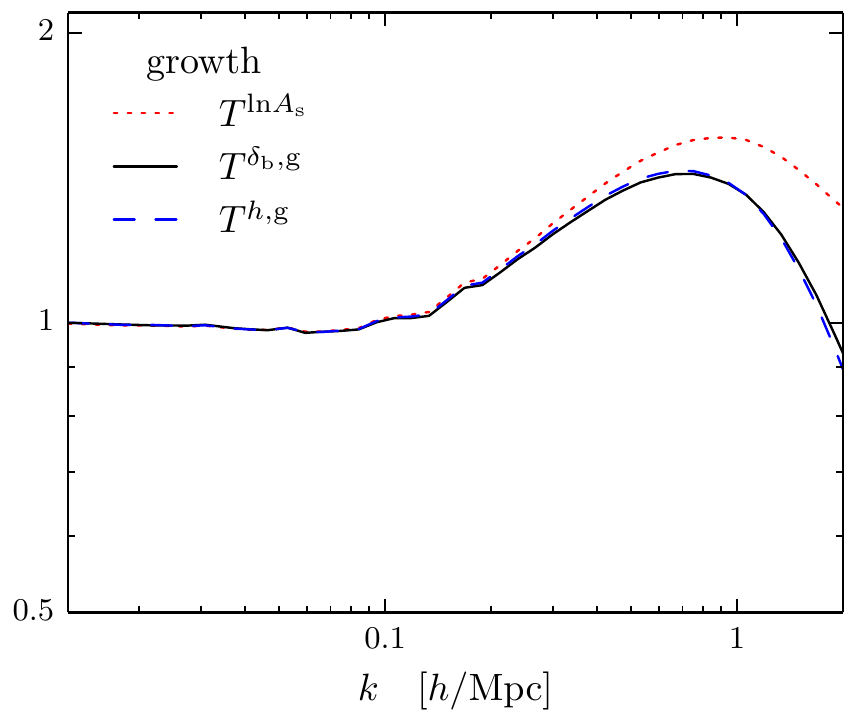}
    \caption{
        \footnotesize Growth component of the power spectrum response to $\delta_\br$ and
        $h$ compared with the response to the initial amplitude of power $\lnAs$.
        The two growth responses are nearly identical  whereas the $\lnAs$ one differs in the
        nonlinear regime.  
       The difference can be attributed to a change in halo scale radii between models
       with the same linear power spectrum. 
    }
    \label{fig:growth}
\end{figure}

\begin{figure}[tb]
    \centering
    \includegraphics[width=3.4in]{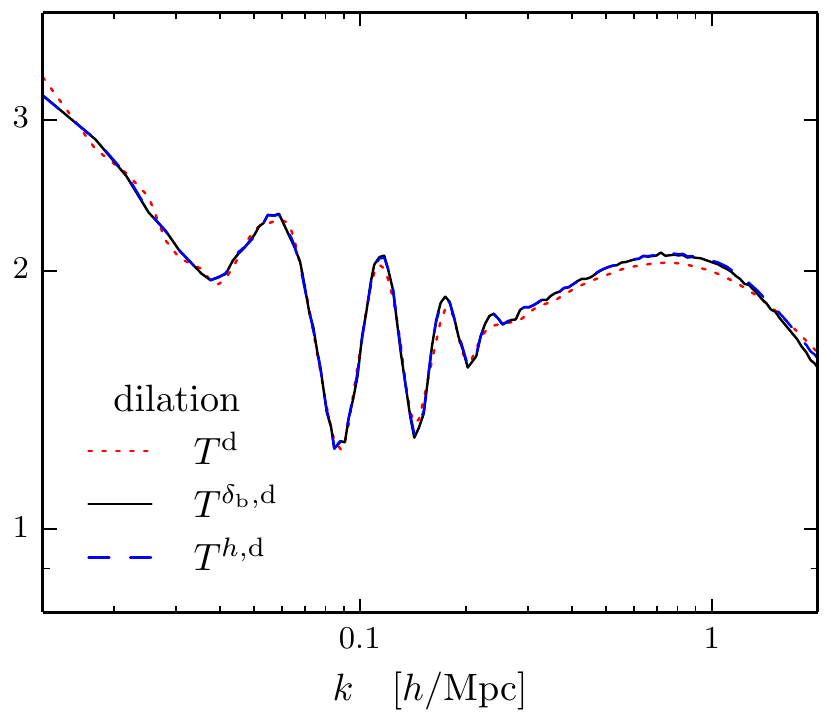}
    \caption{
        \footnotesize Dilation component of the power spectrum response to $\delta_\br$ and
        $h$ compared with that derived from the slope of the power spectrum $T^\textrm{d}$.   All three responses are nearly identical with the $T^\textrm{d}$ exhibiting more noise and resolution
        dependence from numerical differencing. 
    }
    \label{fig:dilation}
\end{figure}

The response to $h$ shares the same features as $\delta_\br$.   
Since $h$ is varied at fixed $\Omegab h^2$ and $\Omegac h^2$ in a flat universe
its impact on
the power spectrum in the linear regime
 at a fixed scale in Mpc comes solely from changing the growth function $D$ due to the change in 
 $\Omega_\Lambda$.   However because
 observations at $z=0$ determine a scale in \hMpci, observable features in the power spectrum shift.  This is the same effect that allows BAO to measure $h$ at $z=0$ or
 the expansion rate and angular diameter distance at higher redshift.
 The result is that the response can be decomposed as 
 \begin{equation}
    T^h = 2\frac{\partial\ln D}{\partial h}T^{h,\textrm{g}}  + \frac{1}{h}T^{h,\textrm{d}}.
    \label{eq:Th}
\end{equation}
For reference, in the chosen cosmology $\partial \ln D/\partial h \approx -0.668$.
As shown in  Fig.~\ref{fig:growth}, the growth pieces of the $\delta_\br$ and $h$ responses
are nearly indistinguishable.   Likewise, defining the dilation piece as the difference
of these responses agrees with the dilation defined from $\delta_\br$
and 
the $k$-derivative
of the mean power spectrum
as shown in Fig.~\ref{fig:dilation}.
In an extended parameter space, 
 we expect the response to parameters 
 such as
the dark energy equation of state or curvature can be modeled accurately with
$T^{h,\textrm{g}}$ and their impact on the linear growth function $D$.

\begin{figure*}[tb]
    \centering
    \includegraphics[width=3.4in]{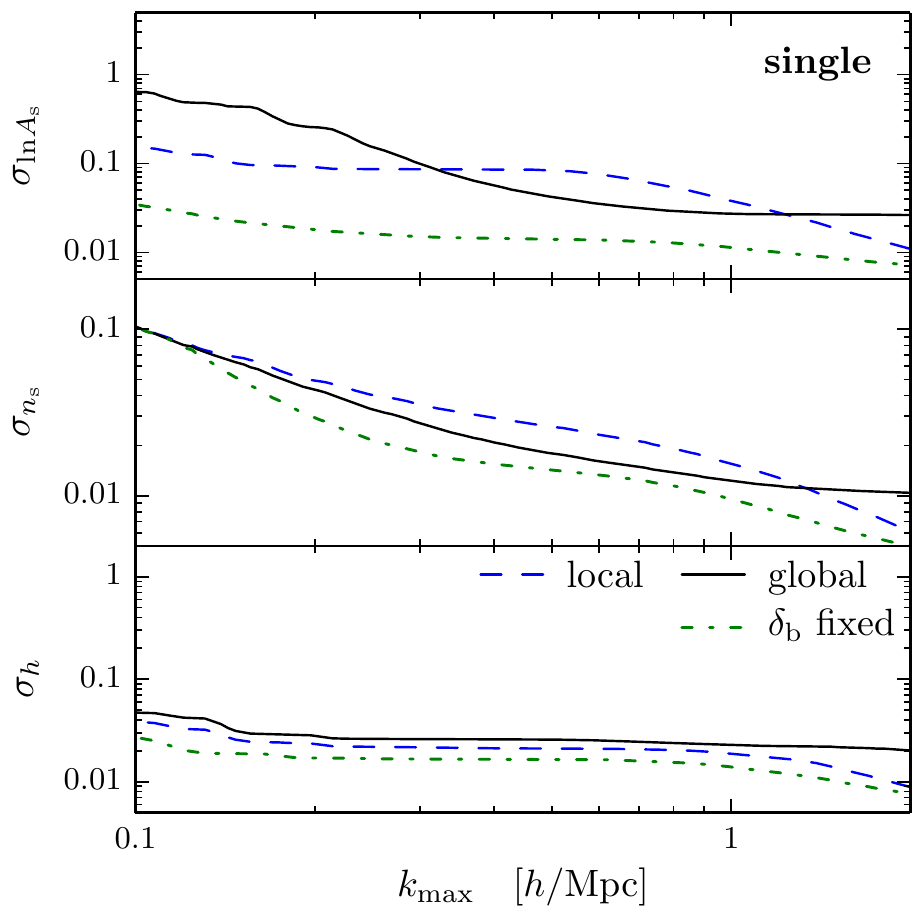}
     \includegraphics[width=3.4in]{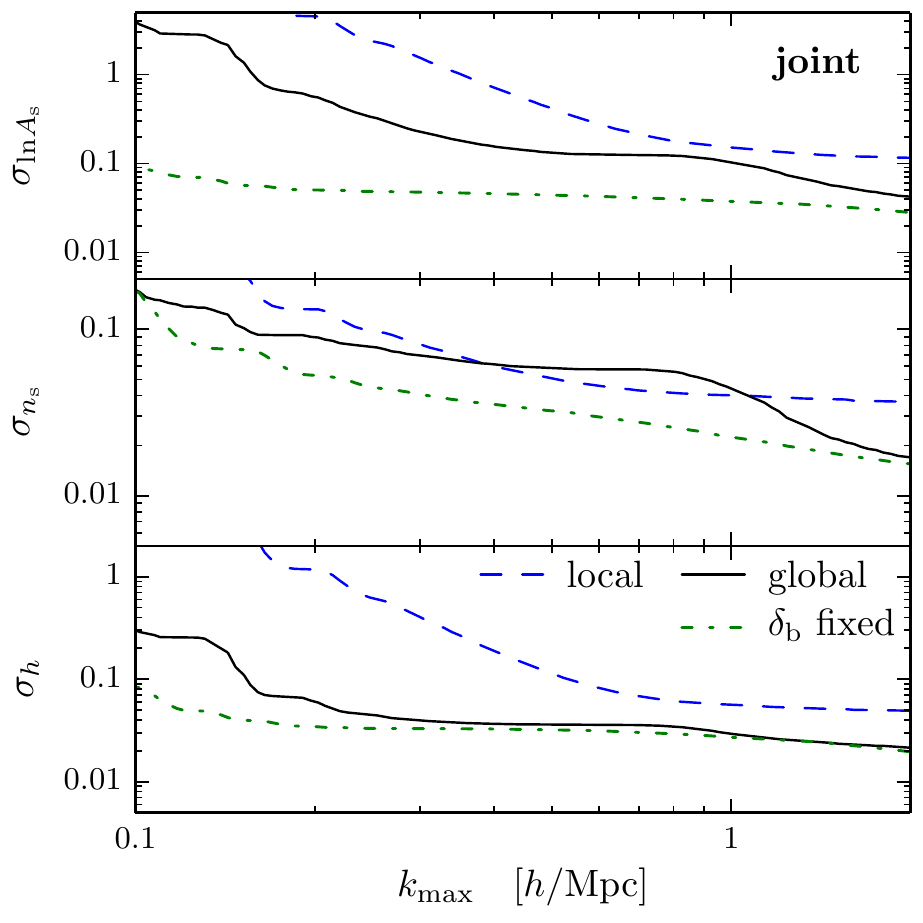}
    \caption{
        \footnotesize  Errors on cosmological parameters as a function of the maximum 
        $k$-bin for $\delta_\br$ fixed (dot-dashed), marginalized in the local case
        (dashed) and in the global case (solid).  Left panel: single cosmological parameter
        estimation with the remaining two held fixed.
        Right panel: joint cosmological parameter estimation.  
    }
    \label{fig:kmaxcosmo}
\end{figure*}

Next $\lnAs$ directly controls the amount of  power in the spectrum at the initial epoch.   At $z=0$ a change in $\lnAs$
and a change in growth can produce the same linear power spectrum and hence in the linear regime the
two responses are indistinguishable.      In Fig.~\ref{fig:growth}
we compare these response and show that they begin to differ in the
nonlinear regime.    If the nonlinear power spectrum
were a 
functional 
of the linear power spectrum then the responses would be identical. 
For example, in a halo model description, if the mass function were universal and the
scale radii of halos {were}
fixed then the linear power spectrum determines the nonlinear 
power spectrum directly with no further reference to the past history of structure formation. 
We have verified that  the differences in response are qualitatively modeled by a 
change in concentration of halos with respect to a fixed scale \footnote{Fixed as opposed to a parameter dependent scale such as the virial scale.  For example $r_{180}$ where
the spherical overdensity is 180 times the mean matter density.}.
The concentration of halos retains information about the 
{mean}
density of the Universe at
their formation epoch, 
and so one would not expect a change in the initial conditions
and late-time growth that leaves the linear power invariant  to yield the same concentration.
In principle, then $\lnAs$ is not degenerate with other parameters associated with growth.
In practice,  uncertainty in the concentration-mass relation due to baryonic physics
can restore this degeneracy.

Small changes in the tilt can also restore this degeneracy across a limited range in $k$.  
Fig.~\ref{fig:Tp} also shows the response to tilt, which changes the
power spectrum in opposite directions around the pivot point $k=0.05~${Mpc}$^{-1}=0.071\,$\hMpci.  Given the greater statistical power of measurements in the nonlinear regime, a small
amount of tilt can compensate the differences due to changes in the concentration.

In summary, the response functions show that for power spectrum measurements
with respect to the local mean, there are 3 parameters $\{ \delta_\br, h, \lnAs \}$ whose power 
spectrum response is  characterized mainly by linear combinations of two templates, $T^\textrm{g}$
and $T^\textrm{d}$ once minor changes in tilt or {halo} concentration are
factored in. We therefore expect a strong degeneracy in the local case.  For power
spectra measured with respect to the global mean, the addition of 2 to the response breaks this degeneracy yielding three templates for
three parameters.  
We shall now see that these features are reflected in the forecasted
parameter errors.

\subsection{Parameter constraints without $\delta_\br$ prior}
\label{sub:degen}

We begin with parameter constraints for the case where there is no
external prior on $\delta_\br$ so that any information about it  must be recovered
from the power spectrum measured by the survey.
Fig.~\ref{fig:kmaxcosmo} shows an overview of the impact of marginalizing $\delta_\br$ on
cosmological parameter estimation.
The left panel shows the impact on the 3 cosmological parameters considered one at
a time with the other 2 fixed whereas the right panel shows the result with the other two
marginalized.    In the former case, the degradation 
{is} 
the most severe when
the two templates are most similar.   For example in the global case,  the flattening of
the response in the nonlinear regime relative to local in Fig.~\ref{fig:dPddT} makes
it more similar to $\lnAs$ around $k\sim 1-2\,$\hMpci\ and causes a larger
degradation in errors for such choices of $k_{\rm max}$.  
Parameter degeneracies also increase the importance of having a sufficiently
large $k_{\rm max}$ to distinguish the responses compared with a naive
quantification of information through $F_{\mu\mu}$
\cite{RimesHamilton:05,RimesHamilton:06} (see also \cite{TakadaJain:09,Satoetal:09}).

Similar statements apply for the case where the two other cosmological parameters
are also marginalized.   Here what is important for determining degeneracies is whether
a  linear combination of the four responses can compensate each other.   
From Fig.~\ref{fig:kmaxcosmo} (right) we see that the flattening of the global vs.~local response
now has the opposite effect on $\lnAs$.  By $k_{\rm max}=2\,$\hMpci\ most 
degeneracies are broken for the global case whereas they remain strong and impact
all three cosmological parameters  in the local case.

\begin{figure*}[htbp]
    \centering
    \includegraphics[width=3.5in]{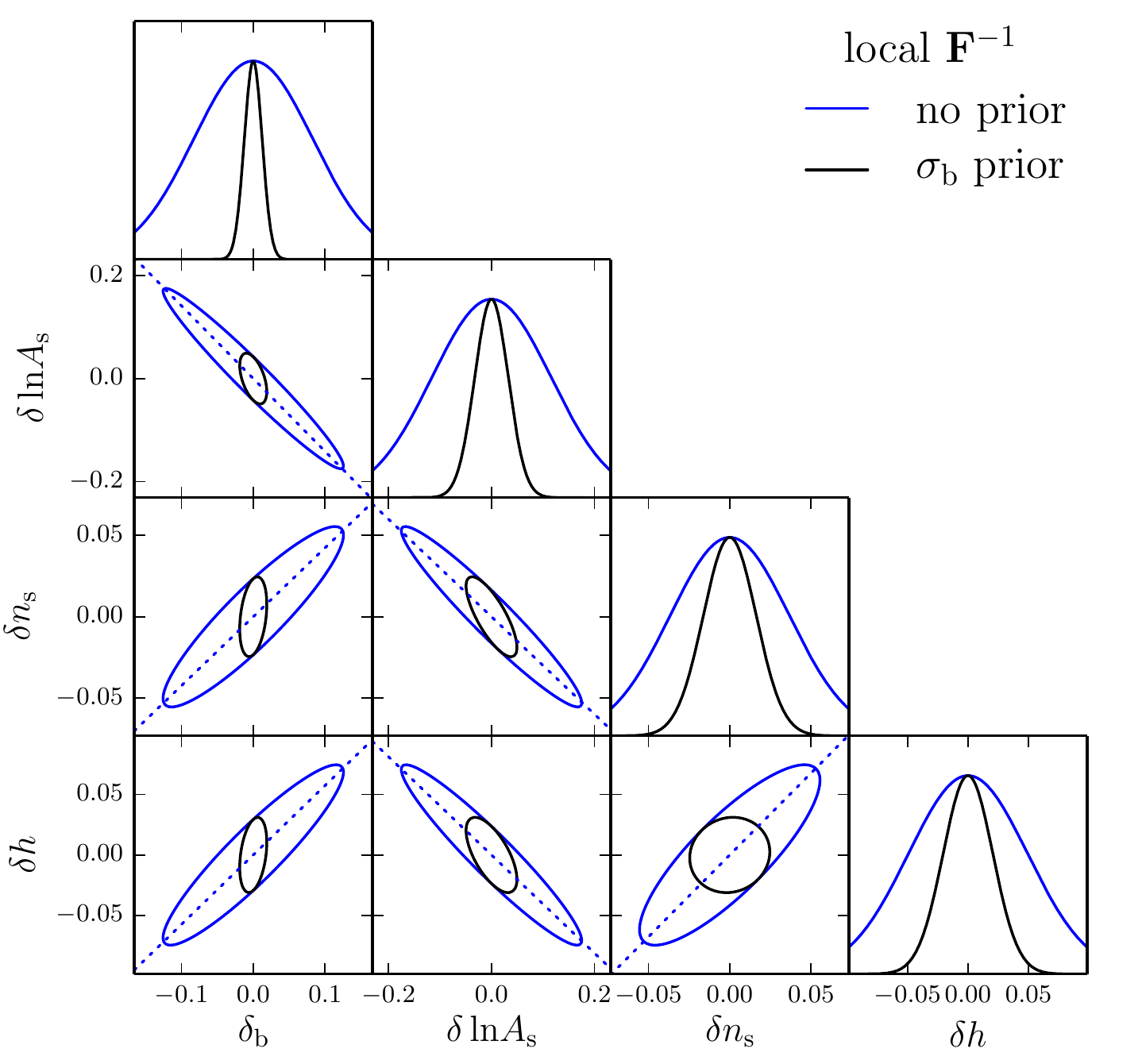}
    \includegraphics[width=3.5in]{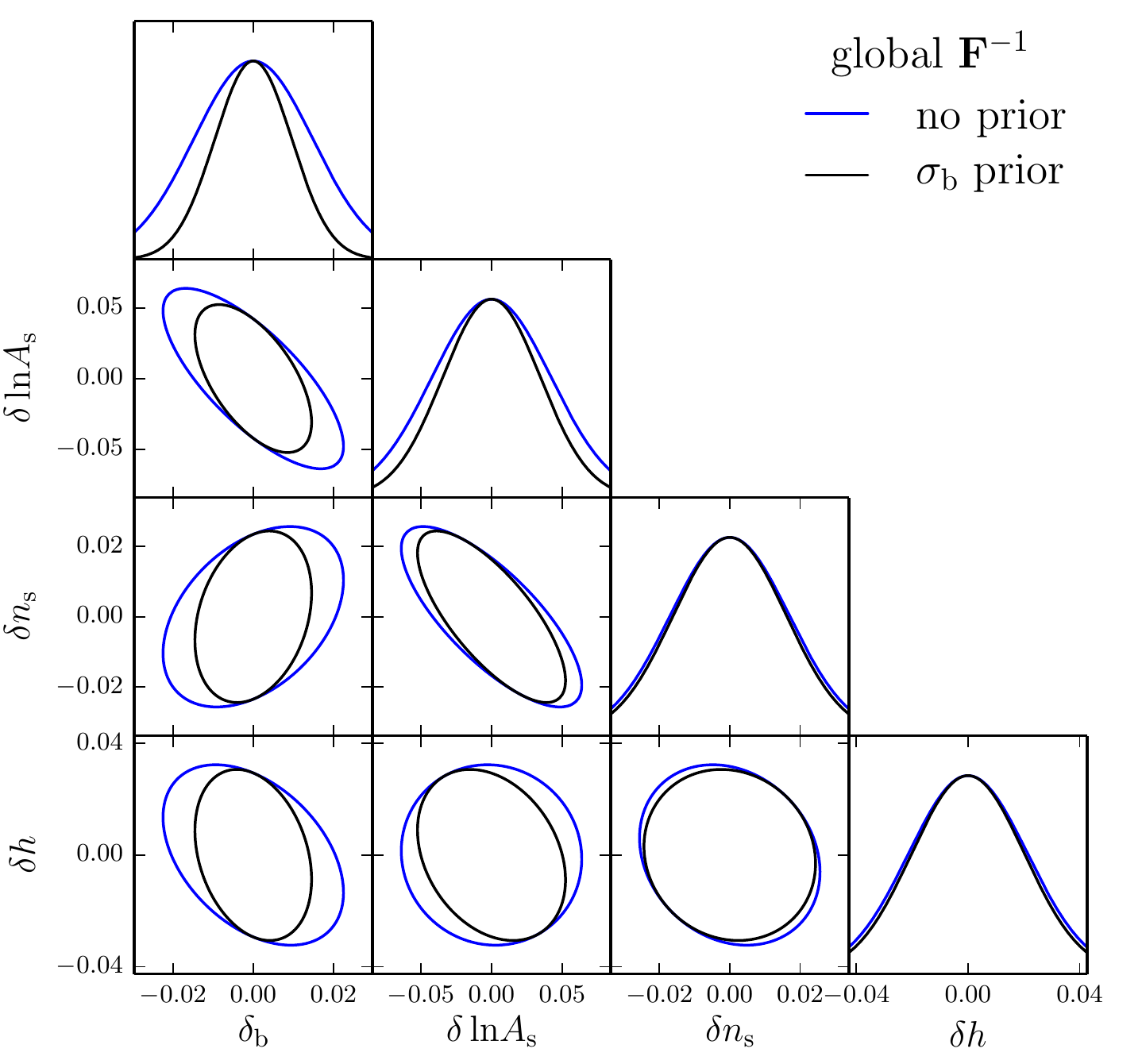}
    \caption{
        \footnotesize Error contours (68\%) and distributions for joint $\delta_\br$ and
        cosmological parameter estimation with $k_{\rm max}=2\,$\hMpci\ with (inner, black) and without (outer, blue) an external prior from $\sigma_\br$.  Left panel: local case,
        with dashed lines representing the degeneracy direction from Eq.~(\ref{eq:localdegeneracy}).
        Right panel: global case.  
}
    \label{fig:ee}
\end{figure*}

To see how these results arise geometrically,
 we show various 2D 68\% confidence regions for the 4 parameters for the local case 
    in Fig.~\ref{fig:ee} (left) with $k_{\rm max}=2\,$\hMpci.   In particular, the contours
{involving}
$\delta_\br$ show a strong degeneracy with all three of the other parameters.
This implies that without a prior on $\delta_\br$, constraints on cosmological parameters
will be severely degraded by marginalizing $\delta_\br$.
This degradation is quantified in Tab.~\ref{tab:degrade} (see Appendix \ref{sec:errordeg}
for details and notation).
With the Fisher matrix, we can use the eigenvector with the  largest
variance
 to identify
the degenerate direction 
\begin{equation}
    \pi_{\rm degen} = 3.5\,\delta_\br + 2.0\,\delta h - 4.9\,\delta\lnAs + 1.5\,\delta\ns,
    \label{eq:localdegeneracy}
\end{equation}
where we have normalized the vector
such that the estimator of this mode has unit variance.
This direction is displayed as the dashed lines in Fig.~\ref{fig:ee} (left).  

Fig.~\ref{fig:Tdegen} shows the response of the power spectrum to this combination.
This linear combination nulls the response in the nonlinear region where changes in the tilt
and {halo} concentration can compensate each other.  The change in tilt implies a
change in the linear regime which variations in $h$ partially compensate 
 at the expense of leaving a residual response in the
BAO scale.

\begin{table}[tb]
    \centering
    \begin{tabular}{@{\hspace{2.5em}}c@{\hspace{1.5em}}c@{\hspace{1em}}c@{\hspace{1.5em}}c@{\hspace{1em}}c@{\hspace{.5em}}}
        \hline 
        & \multicolumn{2}{c}{local}\hspace{1em} & \multicolumn{2}{c}{global} \\
        parameter &  no prior   & $\sigma_\br$  prior  & no prior     & $\sigma_\br$  prior \\\hline
        \multicolumn{1}{l}{$V=V_0$} &&&&
         \\ 
        $\lnAs$
        & 4.14  & 1.17  & 1.51  & 1.24 \\
        $\ns$
        & 2.37  & 1.05  & 1.09  & 1.04 \\
        $h$
        & 2.54  & 1.06  & 1.10  & 1.04 \\
        max,$\delta_\br$
        & 5.02  & 1.24  & 4.34  & 2.91 \\\hline
          \multicolumn{1}{l}{$V=100V_0$} &&&&
         \\        
         $\lnAs$
         & 4.14  & 1.09  & 1.51  & 1.15 \\
        $\ns$
        & 2.37  & 1.03  & 1.09  & 1.03 \\
        $h$
        & 2.54  & 1.03  & 1.10  & 1.03 \\
        max,$\delta_\br$
        & 5.02  & 1.13  & 4.34  & 2.37 \\
        \hline
    \end{tabular}
    \caption{Error degradation from marginalizing $\delta_\br$, with or without prior,
     for the local and global cases, the fiducial $V_0$  
     and $100V_0$ volumes and
         $k_\textrm{max}=2\,$\hMpci.  Note that ``max'' is the combination of cosmological parameters that is maximally degraded by marginalizing $\delta_\br$ and that degradation is
         numerically equal to that on the errors of $\delta_\br$ by marginalizing cosmological
         parameters.
    }
    \label{tab:degrade}
\end{table}

Next in Fig.~\ref{fig:ee} (right) we show the 2D error contours for
the global case.   As discussed in the previous
section, the change in the $\delta_\br$ response from the local case eliminates
the near degeneracy provided by the growth and dilation responses.
Consequently the degradation in parameter errors from marginalizing $\delta_\br$ 
is much smaller as shown in Tab.~\ref{tab:degrade}.    The largest degradation occurs for
$\lnAs$ and involves mainly the direction 
\begin{equation}
    \pi_{\rm degen} = 5.5\,\delta_\br - 0.22\,\delta h - 20\,\delta\lnAs + 6.4\,\delta\ns.
    \label{eq:globaldegeneracy}
\end{equation}
Fig.~\ref{fig:Tdegen} shows that this direction does not allow  an approximate nulling of the response in the nonlinear region.  Instead it 
 yields a flat response
in the nonlinear region which no longer requires significant variations in $h$ to
compensate in the linear regime.   

Although these degeneracy 
studies, 
which identify the worst constrained directions, 
reveal the overall impact on the individual cosmological parameters, it is also
interesting to consider the impact of marginalizing $\delta_\br$ on the best
constrained directions.    For example, if the CMB or other cosmological probe
constrains a different combination
of these parameters and breaks the degeneracy, then the impact of the best
constrained directions may be revealed.    
Moreover, the impact of marginalizing
$\delta_\br$ is generally larger on the best rather than worst constrained directions. 
In Appendix \ref{sec:errordeg}, we formalize these statements by identifying
the combination of cosmological parameters whose errors are most degraded
by marginalizing $\delta_\br$,
\begin{eqnarray}
    \pi_{\rm max} &=& 127\,\delta h + 138\,\delta\lnAs + 183\,\delta\ns, \;\; \textrm{global},
    \nonumber \\
    \pi_{\rm max} &=& 67.9\,\delta h + 106\,\delta\lnAs + 116\,\delta\ns, \;\; \textrm{local},
    \label{eq:max}
\end{eqnarray}
which in practice are in directions very similar to
the best constrained direction without marginalization
\begin{equation}
    \pi_{\rm best} = 125\,\delta h + 137\,\delta\lnAs + 199\,\delta\ns,
    \label{eq:best}
\end{equation}
especially in the global case.
Here again we normalize each mode so that its estimator has unity variance,
conditional on $\delta_\br$ held fixed.
The amount of the
degradation is numerically equal to that on $\delta_\br$ upon marginalizing over
cosmological parameters.    In Tab.~\ref{tab:degrade} we compare the various degradation
in errors.   Note that even in the global case, the maximal degradation is large
and comparable to the 
{local}
 case.

Finally given that the covariance matrix scales with survey volume according
to Eq.~(\ref{eq:CijV}), we can account for a change in survey volume by
simply rescaling all parameter errors according to Eq.~(\ref{eq:sigmaV}) so that the
relative impact of $\delta_{\br}$ remains the same.

In summary, even though the local case involves a smaller response to $\delta_\br$,
it can have a much bigger impact on cosmological parameter estimation compared with the global
case  due to the ability to construct
near perfect degeneracies in the nonlinear regime.   On the other hand, errors in 
the combination of cosmological parameters along the direction that is best constrained without $\delta_\br$
are substantially degraded in both cases.

\begin{figure}[tb]
    \centering
    \includegraphics[width=3.4in]{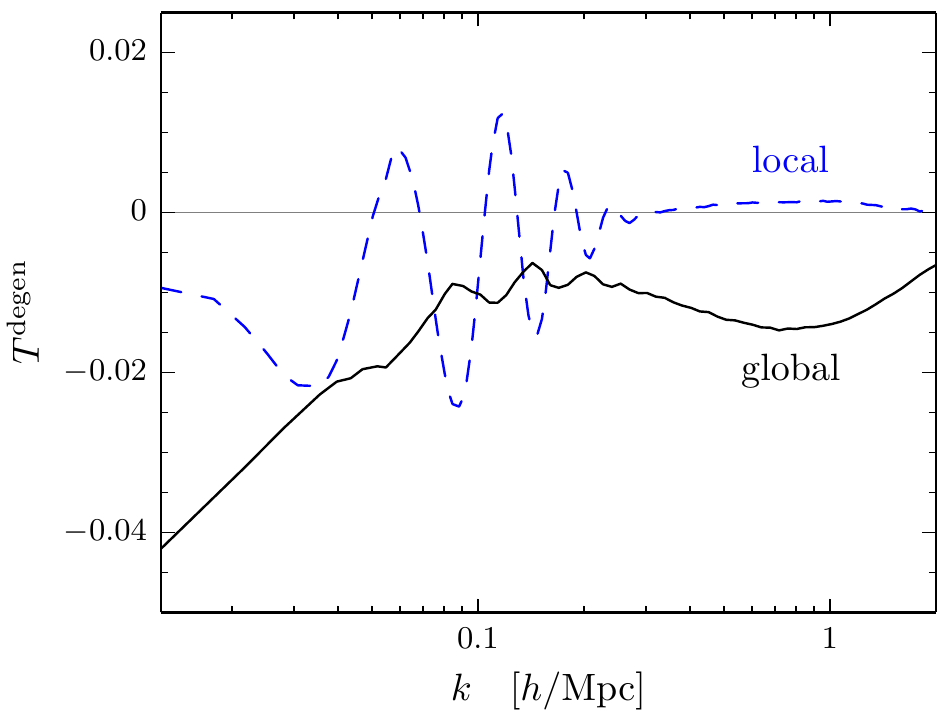}
    \caption{
        \footnotesize Power spectrum responses in most degenerate directions in the
        local and global cases, Eq.~(\ref{eq:localdegeneracy}) and (\ref{eq:globaldegeneracy}), respectively.  In the local case, the direction effectively nulls the response in the
        nonlinear regime creating a near perfect degeneracy.  In the global case, no direction
        nulls the response and the degeneracy is much weaker.   Responses are normalized
        to represent $1\sigma$ deviations in degenerate parameter combination.
    }
    \label{fig:Tdegen}
\end{figure}

\subsection{Parameter constraints with $\delta_\br$ prior}
\label{sub:degen_prior}

If the linear power spectrum were perfectly predicted by external information on 
cosmological parameters {such as the CMB measurements,}
%, 
then we would possess prior knowledge that $\delta_\br$ is 
distributed as a Gaussian with variance $\sigma_\br^2$.  In this limiting case
\begin{equation}
    F_{\mu\nu}^{\rm prior} = \frac{1}{\sigma_\br^2}
    \delta^\mathrm{K}_{\delta_\br \mu}\delta^\mathrm{K}_{\delta_\br \nu},
\end{equation}
where $\delta^\mathrm{K}_{\mu\nu}$ is the Kronecker delta.
Here we use Eq.~(\ref{eq:sigma}) to evaluate
$\sigma_\br$.
In practice, cosmological parameter uncertainties should be propagated into this
prior but this simple prior is useful to study as a best case scenario.   Note that
the view of super-sample effects as signal makes it simple to incorporate uncertainties
in $\sigma_\br$ as opposed to the excess covariance approach where it would enter through
the cosmological parameter dependence of the covariance matrix.

The impact of this prior is qualitatively different in the local and global cases.
In the local case, with all 4 parameters jointly estimated,
intrinsic parameter degeneracies are so strong that the net
degradation in parameter estimation from marginalizing $\delta_\br$ is prior limited
as shown in Fig.~\ref{fig:ee} (left) and Tab.~\ref{tab:degrade}.    For the global case, the degradation is
only marginally changed by the prior.   
In both cases, with the prior, the remaining impact of marginalizing $\delta_\br$ is mainly on
$\lnAs$ and represents approximately a factor of $1.2$ degradation in its errors.

The reason that these degradations are small in both cases is the remaining degeneracies
among the cosmological parameters themselves come to dominate the parameter errors.
This fact however does not mean that marginalizing $\delta_\br$ has little effect
on combination of cosmological parameters that are well constrained in its absence.   
The direction of maximal degradation is the same with or  without the $\sigma_\br$ prior
and so is close to the best constrained direction in Eq.~(\ref{eq:best}).   In the
global case, this degradation is nearly a factor of 3 and is substantially larger than
the local case.   This degradation can be important for errors on individual cosmological
parameters if other measurements break the intrinsic cosmological
parameter degeneracies.

Finally, with a $\delta_\br$ prior, parameter errors do not simply scale with volume as
$V^{-1/2}$ unless $\sigma_\br \propto V^{-1/2}$.    To test the impact of volume scaling
we consider a case where the volume is increased by $V/V_0=100$.   In this case
$\sigma_\br V^{1/2}$ is reduced by a factor of $1.4$.
%(commented out since the ratio table was removed)
%The ratios of the degradation factors corresponding to Tab.~\ref{tab:degrade}.
For cases that are not
limited by the prior, the volume scaling of $C_{ij}$ ensures that the degradation factors
are the same.
%, i.e.~a ratio of unity.
In general for $\Lambda$CDM, 
increasing the volume tends to slightly
strengthen the relative impact of the prior.

\section{Discussion}
\label{sec:discussion}

Super-sample density fluctuations systematically change the observable
 power spectrum of sub-sample modes.
In this paper, we have developed and tested the interpretation of this effect
as a signal due to the dependence of the observed power spectrum on the
mean density fluctuation in the survey volume $\delta_\br$.  This dependence
can be calibrated efficiently by using the separate universe technique that absorbs the
fluctuation into a change in the background parameters.  

This interpretation has the advantage that the effect can be incorporated
into parameter estimation without modification of traditional procedures or
pipelines.   The form of the likelihood of the power spectrum
data as a function of the model takes the same form except that the model
gains a parameter $\delta_\br$ in addition to cosmological parameters.  Its 
impact on cosmological parameter estimation comes through parameter 
degeneracies.
Contrast this with the alternate but equivalent view that when ensemble averaged over
many realizations of the survey volume and $\delta_\br$, the effect induces
a covariance in the power spectrum that modifies the form of the likelihood
function. 
 
The super-sample signal allows $\delta_\br$ itself to be estimated from power
spectrum data.    The amount of information on $\delta_\br$ depends on whether
the power spectrum is measured with respect to the global or
local mean density, which is relevant, e.g., for weak lensing or galaxy
clustering, respectively.   For a wide range of survey volumes, the global case
contains substantial extra information in the nonlinear regime on top of 
the prior expectation that it be limited by the rms $\sigma_\br$ predicted
by linear theory.   For the local case, the extra information is comparable 
to the prior for $k_{\rm max}=2\,$\hMpci. 

If cosmological parameters are jointly estimated, this extra information can be
lost to degeneracies.   Likewise, marginalization of $\delta_\br$ can degrade
errors on cosmological parameters.   
The degradation
takes on different values depending on the
 the parameter space and priors considered, whether the local or global
power spectra are used, and the maximum wavenumber 
utilized.
In general, the strongest degeneracies arise from compensated
changes in
the growth of structure and dilation of features induced by the parameters, which provides
a physical basis for which to extend our results beyond the $\Lambda$CDM parameter space.  
  Without 
prior information on $\sigma_\br$, the degradation of errors for the local case can be more than a factor of 4 for $k_{\rm max}=2\,$\hMpci\ and likewise in the global case but only for the combination of
cosmological parameters that is best measured in the absence of $\delta_\br$ and maximally degraded by its marginalization.
  Even with a prior
that reflects perfect knowledge of the linear power spectrum, the maximal degradation in the
global case can reach a factor of 2--3.  Fortunately all of these cases and more can be simply and rapidly
considered given a single calibration of the power spectrum responses.

While we have only considered the effects on the matter power spectrum through
$N$-body simulations in $\Lambda$CDM, our separate universe and growth-dilation techniques can be extended to other parameter spaces or observables
and can incorporate baryonic effects and galaxy formation. While other uncertainties in modeling the data, e.g.\ redshift space distortions and galaxy bias for galaxy surveys,
may  dominate the error budget for current measurements, the super-sample mode provides
an intrinsic limitation to extracting cosmological information that is degenerate with its effects.

\vfill

\smallskip{\em Acknowledgments.--}  
We thank M.\ Becker, M.\ Busha, B.\ Erickson,  G.\ Evrard, N.\ Gnedin,  A.\ Kravtsov,
D.\ Rudd,  R.\ Wechsler,  and the University of Chicago
Research Computing Center   for running, storing, and
allowing us to use the large-volume simulations in this study. 
{We also thank M.\ Becker, D.\ Rudd., N.\ Gnedin, A.\ Kravtsov and F.\ Schmidt for useful discussions.}
YL and WH  were supported
 by U.S.~Dept.\ of Energy
 contract DE-FG02-13ER41958
 and the
 Kavli Institute for Cosmological Physics at the University of
 Chicago through grants NSF PHY-0114422 and NSF PHY-0551142.   MT was
 supported by World Premier International Research Center Initiative
 (WPI Initiative), MEXT, Japan, by the FIRST program “Subaru
 Measurements of Images and Redshifts (SuMIRe)”, CSTP, Japan, and by
 Grant-in-Aid for Scientific Research from the JSPS Promotion of Science
 (Nos. 23340061 and 26610058). 
This work was also supported in part by the National Science Foundation
under Grant No. PHYS-1066293 and the hospitality of the Aspen Center for
Physics, where this work was completed. 
Plots in this work are made with \texttt{matplotlib}~\cite{Hunter:2007}.

\vfill

\appendix
\section{Numerical Implementation}
\label{app:sim}

In this Appendix we provide details on various numerical calculations in the main paper.
In \S~\ref{sub:sims}, we describe the cosmological simulations used in the super-sample
signal studies.  We calibrate the power spectrum response to various
parameters in \S~\ref{sec:response}.   In \S~\ref{sec:errordeg}, we establish the formalism for
calculating the degradation of parameter errors upon marginalizing over $\delta_\br$.

\subsection{Simulations}
\label{sub:sims}

\begin{table}[tb]
    \centering
    \begin{tabular}{@{\hspace{.5em}}c@{\hspace{1em}}c@{\hspace{1em}}c@{\hspace{1em}}c@{\hspace{1em}}c@{\hspace{.5em}}}
        \hline
        $\Omegam$ & $\Omegab$ & $h$ & $n_\textrm{s}$ & $\sigma_8$ \\
        \hline
        0.286 & 0.047 & 0.7 & 0.96 & 0.82 \\
        \hline
    \end{tabular}
    \caption{\footnotesize Parameters of baseline flat $\Lambda$CDM model used throughout.}
    \label{tab:LCDMpar}
\end{table}

Here we summarize the salient features of the simulations and power spectrum analysis from 
Ref.~\cite{Lietal:14} and  
used in \S~\ref{sec:sss} to test super-sample effects with subvolumes of
%a 
large-volume simulations. 
For the large-volume simulations, we take a suite of 
7 realizations of the fiducial cosmology given 
%by 
{in Tab.~\ref{tab:LCDMpar},} 
originally made for the Dark Energy Survey.
Each of these has a $4\,h^{-1}\textrm{Gpc}$ box length evolved from initial conditions at
 $a_\textrm{i}=0.02$
%.
{that are}
provided by \texttt{CAMB} \cite{Lewis:1999bs,Howlett:2012mh}
and \texttt{2LPTIC} \footnote{\href{http://cosmo.nyu.edu/roman/2LPT/}{http://cosmo.nyu.edu/roman/2LPT/}},
using \texttt{L-Gadget2} \cite{Springel:2005nw} with
$2048^3$ particles and $3072^3$ (Tree-)PM 
grid.  
We then assign the particles to 
a 
$(8 \times 1920)^3$ 
grid 
with a cloud-in-cell (CIC) scheme,
before subdividing each large box into $8^3=512$ subvolumes of size 
\begin{equation}
V_0 = (500 h^{-1} \textrm{Mpc})^3
\end{equation}
for a total of $N_\textrm{s}=3584$ subboxes.  

In each subvolume, we extract the mean density fluctuation $\delta_\br$
and 
the power spectrum %of the subboxes 
by FFT.   For the power spectrum, we then 
deconvolve the CIC window
and bin the result to 80 logarithmically spaced $k$-bins per decade to form 
$\hat \Dv^\textrm{sub}_i$. 
Each bin is positioned at the average $k_i$ weighted by the number of modes.
This binning scheme is used throughout the paper.

To calibrate the mean power spectrum $\Dv_i$ and its covariance matrix $C_{ij}$, in
the absence of $\delta_\br$, 
we use the same number $N_\textrm{s}$ of simulations
of 
the same size as the subboxes but with $256^3$ particles and $512^3$
(Tree-)PM grid.
{We measure the power}
spectrum $\hat \Dv^\textrm{sm}$ of each small-box simulation
in the same way with a $1920^3$ grid. 
All {the numerical}
settings match the $1/8$ scaling of the large box dimensions except
for the (Tree-)PM grid.

The mean power spectrum of the subboxes differs from that of the small boxes in two ways.
On large scales the difference is dominated by  convolution bias from the subbox window,
and on small scales there is a $1\%$ difference due mainly to
using different resolutions for the (Tree-)PM grid in simulations.
We debias the estimator as in Ref \cite{Lietal:14} by rescaling
\begin{equation}
    \hat \Dv_i = \frac{\Dv^\textrm{sm}_i}{\Dv^\textrm{sub}_i}  \hat \Dv^\textrm{sub}_i,
    \label{eq:debias}
\end{equation}
where we have defined 
\begin{equation}
    \Dv^\textrm{X}_i \equiv \frac{1}{N_\textrm{s}} \sum_{a=1}^{N_\textrm{s}}  \hat \Dv_i^{\textrm{X},a}
\end{equation}
as the average over the $N_\textrm{s}$ samples.  Thus 
\begin{equation}
    \langle \hat \Dv_i \rangle =  \Dv^\textrm{sm}_i \equiv  \Dv_i.
\end{equation}
For the power spectra referenced to the subbox mean
\begin{equation}
    \hat \Dv_i^W = \frac{\hat \Dv_i}{(1+\delta_\br)^2},
\end{equation}
where $\delta_\br$ is the average density fluctuation in the same box.

{Next we estimate}
%for 
the covariance matrix in the absence of $\delta_\br$ 
{as}
\begin{equation}
    C_{ij} = \frac{N_\textrm{s}}{N_\textrm{s}-1}
    \Bigg[ \frac{\sum_{a=1}^{N_\textrm{s}} \hat \Dv^{\textrm{sm}, a}_i \hat \Dv^{\textrm{sm},a}_j}{N_\textrm{s}}
    - \Dv^\textrm{sm}_i\Dv^\textrm{sm}_j  \Bigg].
\end{equation}

Finally unless otherwise specified, the simulations used to construct the
response functions in the next section follow the prescription for 
the  small  box simulations in order to preserve the same mass and force resolution.

\subsection{Response calibrations}
\label{sec:response}

In this section we provide some detail on the calibration of the power spectrum response
to $\delta_\br$ and the cosmological parameters.   We review the $\delta_\br$ response
calibrated in Ref.~\cite{Lietal:14} and illustrate the cosmological parameter calibration with
$h$ as it demonstrate all the important concept and techniques.

\subsubsection{$\delta_\br$ response}
\label{sub:SU}

Following the separate universe technique {developed in Ref.~\cite{Lietal:14}}
(see also \cite{Sirko:05,Baldauf:2011bh,Gnedin:2011kj}),
a nonzero mean density fluctuation $\delta_\br$ at $z=0$ can be absorbed into the background by 
a redefinition of the cosmological parameters 
compared to those of a global $\Lambda$CDM universe
\begin{equation}
    \frac{\delta\Omegam}{\Omegam}
    \approx \frac{\delta\OmegaL}{\OmegaL}
    %\approx \frac{-\delta\OmegaK}{1-\OmegaK}
    \approx -2 \frac{\delta h}{h}
    \approx \frac{5\Omegam}{3} \frac{\delta_\br}{D_0}
    \label{eq:SU}
\end{equation}
where the $D_0=D(z=0)$ and the  linear growth function is normalized as
\begin{equation}
    \lim_{a\to0} D = a.
\end{equation}
Note that even if the global universe is flat,
the separate universe would have a nonzero spatial curvature $\delta\Omegam+\delta\OmegaL \ne 0$.  Finally, as discussed in \S~\ref{sub:Ts}, the scale factor 
associated with a given value of $a$ in the global cosmology is shifted by
\begin{equation}
    \frac{\delta a}{a} \approx -\frac{1}{3}\frac{D}{D_0}\delta_\br,
\end{equation}
in the separate universe.   For example, at $z=0$ in the global universe, 
{$z_\textrm{L} = \delta_{\br}/3$} in the local universe.

With the separate universe cosmological parameters set,
we conduct $N$-body simulations to calibrate the response of the power spectrum
by finite difference of models
with 
$\delta_\br =\pm0.01$  evaluated at $z_\textrm{L}$.  
Here and below we always difference simulations
with the  same initial seeds to
suppress the stochasticity from sample variance.
Given that the mean density of the separate universe is reinterpreted as
the local
density of the finite survey, power spectra extracted from these simulations
are always referenced to the local mean, $P^W$ in Eq.~(\ref{eq:Plocal}).

To calibrate the growth response $T^{\delta_\br,{\rm g}}$, we fix the simulation box
in comoving Mpc and difference the results from $\pm \delta_\br$.
This procedure only includes the impact of 
$\delta_\br$ on the growth of structure and omits the fact that due to the difference
in redshift, the physical scale associated with a given comoving scale differs by the
dilation factor.    Separately, we also calculate the total response
$T^{\delta_\br}|_{\rm local}$ by instead fixing the physical scale in Mpc of the simulations
at the final redshift with the same initial seeds in box coordinates.   
Finally we average over 64 pairs of realizations
to reduce the remaining stochasticity to a level that is negligible for our purposes,
with standard errors of the mean of a few percent or better, and at sub-percent level
in the nonlinear regime.

To test the precision of our results,
we have employed simulations with twice the mass and PM resolutions,
to verify that at $z=0$ for $k \lesssim 2~ h$/Mpc the responses have converged to several percent or better.
We refer the readers to Ref.~\cite{Lietal:14} for more details of the calibration pipeline.

With $T^{\delta_\br,{\rm g}}$ and $T^{\delta_\br}|_{\rm local}$ calibrated from simulations
we can construct the dilation response from Eq.~(\ref{eq:Tb})
\begin{equation}
     T^{\delta_\br,\textrm{d}} = -3 \, \Big(  T^{\delta_\br} \big|_\textrm{local}
    - 2\frac{\partial\ln D}{\partial\delta_\br}  T^{\delta_\br,\textrm{g}} \Big).
    \label{eq:Tdd}
\end{equation}
We compare this constructed dilation response in Fig.~\ref{fig:dilation} with
the response calibrated directly as the slope of a
cubic spline fitted to
the mean power spectrum of $N_\textrm{s}$ small box simulations
\begin{equation}
    T^\textrm{d} = \frac{\partial \ln\Dv}{\partial \ln k}.
\end{equation}
The differences in fact reflect that the constructed response reduces the stochasticity from sample variance and the sensitivity to systematic changes in the slope from finite 
resolution.   We have verified using higher resolution simulations that the constructed
response is both more precise and more accurate than the slope-based response.

\subsubsection{$h$ response}
\label{sub:calibrations}

As an example of cosmological parameter response calibration, we choose
$h$ here as it demonstrate all the important concept and techniques,
including the growth-dilation split, and the test of the linear approximation.
We start with the baseline cosmology in Tab.~\ref{tab:LCDMpar} and  utilize a 
suite
of 64 simulations from the small box simulations of \S~\ref{sub:sims}.
As in the $\delta_\br$ {calibration}
we then simulate  pairs of 
models with  $h'= h\pm\delta h$ with the same seeds to form a triplet of simulations
at fixed  $\lnAs$, $\ns$, $(\Omegab h^2)'= \Omegab h^2$ and $(\Omegac h^2)'=
\Omegac h^2$
in a flat universe.
We choose $\delta h=0.02$ in accordance with the $68\%$ confidence limits
constrained by Planck and WMAP polarization data.

For the total response $T^h$, we set the comoving size of simulation boxes
in \hiMpc\ to be the same, i.e.
\begin{equation}
    L'=500\,\textrm{Mpc}/h',
\end{equation}
whereas for extracting the growth component, we set the box scale in Mpc to be the
same
\begin{equation}
    L'=500\,\textrm{Mpc}/h=500\frac{h'}{h}\,\textrm{Mpc}/h'.
\end{equation}

\begin{figure}[tb]
    \centering
    \includegraphics[width=3.4in]{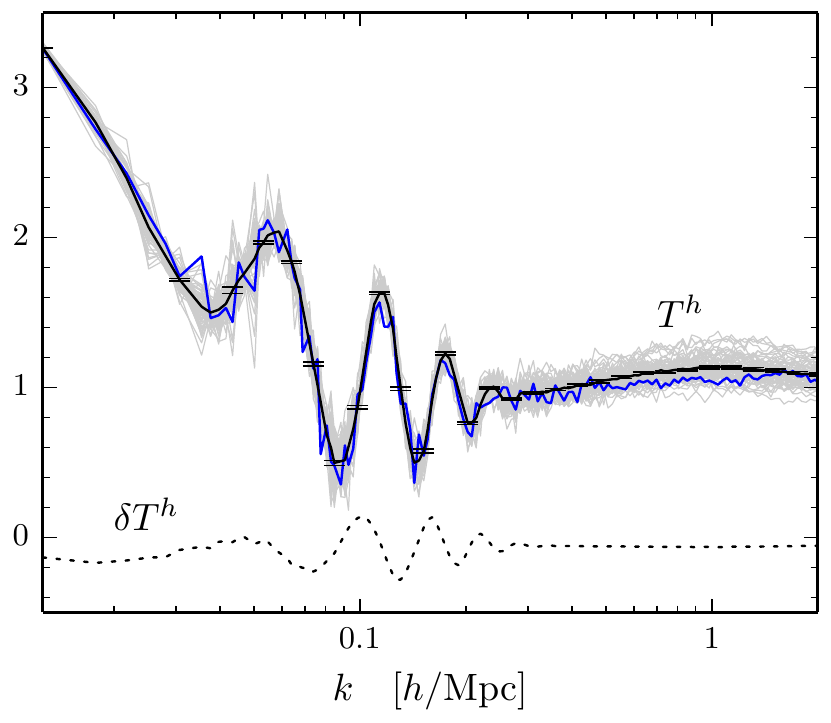}
    \caption{
        \footnotesize Power spectrum response to $h$ from finite differences of
        simulations with $\delta h=\pm0.02$ and scales fixed in \hiMpc.
        Thin gray lines show differences of 64 pairs of realizations,
        with solid black lines representing their means with standard errors.
        A single calibration from one pair of simulations is highlighted in blue
        to illustrate that stochasticity around the mean are highly correlated across nonlinear $k$.
        In the nonlinear regime where most information is located,
        correction from the second derivative in dotted line
        is much smaller compared to the response itself,
        demonstrating that linear response serves as a good approximation for the
        power spectrum variation.
    }
    \label{fig:dPdhT}
\end{figure}

We then difference the binned power spectrum $\hat\Dv$ in box coordinates
in each case to form the required derivative
\begin{equation}
    %\hat T^\mu_i \approx \frac{\ln\hat\Dv_i(+\delta p_\mu)-\ln\hat\Dv_i(-\delta p_\mu)}{2\delta p_\mu},
    \hat T^h  \approx \frac{\ln\hat\Dv(+\delta h)-\ln\hat\Dv(-\delta h)}{2\delta h}.
\end{equation}
  For sufficiently small $\delta h$,
this finite difference converges to the derivative as required for a Fisher matrix calculation.
Each {of the 64 pairs}
%pair in the set of 64 
provides a separate estimate of the response.   In 
Fig.~\ref{fig:dPdhT} we show the individual estimates and the mean of the suite $T^h$.
Note that run-to-run deviations from the mean strongly covary across $k$ in the
nonlinear regime.

Finally, 
to test the linearity of the response, we calibrate the change in the response from the second derivative
\begin{equation}
\delta T^h   =\frac{ \ln\Dv(+\delta h)-2\ln\Dv(0)+\ln\Dv(-\delta h)}{\delta h}.
\end{equation}
using the averages over the 64 triplets.
In Fig.~\ref{fig:dPdhT}, we demonstrate that
%the mean of
this second derivative error term
$|\delta T^h |\ll T^h$.  Specifically, in the fully nonlinear regime where the statistical 
power lies, the error correction is $\lesssim 5\%$.   While we could further reduce
this error by choosing a smaller $\delta h$, this test demonstrates that even for
current uncertainties on this parameter, the Fisher approximation should suffice.

    To test the resolution dependence,
    we have employed $16$ pairs of higher resolution simulations with $512^3$ particles
    and $1024^3$ (Tree-)PM grid to verify that at $k\lesssim2 \,$\hMpci\ 
    our response results have converged to percent level or better.

With both $T^h$ and $T^{h,\rm{g}}$ calibrated in this manner, we can construct
the dilation response using Eq.~(\ref{eq:Th}), 
\begin{equation}
     T^{h,\textrm{d}} = h \, \Big(  T^h - 2\frac{\partial\ln D}{\partial h}  T^{h,\textrm{g}} \Big).
    \label{eq:Thd}
\end{equation}
Similar to the $\delta_\br$ case discussed above, this construction yields a more
accurate and precise dilation response than the slope-based response.

The calibration of $T^\lnAs$ and $T^\ns$ are simpler as no scale dilation is involved,
so that we can use the same box size for simulations.
We take $\delta\lnAs=0.03$ and $\delta\ns=0.01$ again according to the CMB prior,
and the rest of the procedures are the same as for $h$.
For $T^\lnAs$ and $T^\ns$, the linear response assumption is an excellent approximation,
with  second order corrections at the percent level or smaller.

\subsection{Error Degradation}
\label{sec:errordeg}
In the main text, we quote the degradation in the errors in a given cosmological parameter
and the maximal degradation for any linear combination of parameters caused
by marginalizing $\delta_\br$.    Here we give details for those calculations.

The covariance matrix of cosmological parameters with $\delta_\br$ marginalized is
simply the $3\times 3$ subblock of the $4\times 4$ inverse Fisher matrix that contains them.  
We can formalize the extraction of this matrix by defining a $4\times3$ 
projection matrix 
\begin{equation}
{\bf P} = \left(
\begin{array}{c}
 {\bf 0} \\
 {\bf I}_3 \\
\end{array}
\right),
\end{equation}
where ${\bf 0}=(0,0,0)$ and ${\bf I}_3$ is the $3 \times 3$ identity matrix.   Thus
\begin{equation}
    \mathbf{C}_\textrm{mar} \equiv \mathbf{P}^\mathrm{T} \mathbf{F}^{-1} \mathbf{P}.
    \label{eq:Cmar}
\end{equation}
Conversely, if $\delta_\br$ is fixed then the covariance matrix is instead the
inverse of the projected Fisher matrix
\begin{equation}
    \mathbf{C}_\textrm{fix} \equiv [\mathbf{P}^\mathrm{T} \mathbf{F} \mathbf{P}]^{-1}.
    \label{eq:Cfix}
\end{equation}
In the same linear approximation employed in the Fisher analysis, the deviation in any derived parameter from its fiducial value can be thought of
as a linear combination of changes in fundamental parameters,
\begin{equation}
    \pi = \sum_\mu \frac{\partial \pi}{\partial p_{\mu}} \, \delta p_\mu \equiv {\bf v}_\pi \cdot \delta{\bf p},
\end{equation}
where Greek indices in this section run over the 3 cosmological parameters and parameter
vectors ${\bf v}_\pi$ lie in this space.
Thus the degradation in errors $d_\pi$ upon marginalizing $\delta_\br$ is given by
\begin{align}
d_\pi^2 \equiv \frac{\sigma^2_\textrm{mar}(\pi)}{\sigma^2_\textrm{fix}(\pi) } =
\frac{ {\bf v}_\pi^\mathrm{T} \mathbf{C}_\textrm{mar} {\bf v}_\pi }
{ {\bf v}_\pi^\mathrm{T} \mathbf{C}_\textrm{fix} {\bf v}_\pi }.
\end{align}

We can most easily study the degradation in errors in a new basis, called the
Karhunen-Lo\`{e}ve basis.   The 3 basis
vectors ${\bf v}_{k}$ are linearly independent solutions of the generalized eigenvector equation
   \begin{equation}
    \mathbf{C}_\textrm{mar} \mathbf{v}_{k} = \lambda_{k} \mathbf{C}_\textrm{fix} \mathbf{v}_{k}.
      \label{eq:GE}
\end{equation}
Note that $\lambda_{k} = d_{k}^2$, the degradation in the variance of
a parameter defined by $\mathbf{v}_{k}$ and we can normalize these statistically
independent vectors as
\begin{equation}
    {\bf v}_{k}^\mathrm{T} \mathbf{C}_\textrm{mar} {\bf v}_{k'} = \lambda_{k} \delta_{kk'}^{\rm K}, \quad
    {\bf v}_{k}^\mathrm{T} \mathbf{C}_\textrm{fix} {\bf v}_{k'} = \delta_{kk'}^{\rm K}.
\end{equation}
These vectors are not typically orthonormal
in the usual sense ${\bf v}_k \cdot {\bf v}_{k'} \ne \delta_{kk'}^{\rm K}$.

The Karhunen-Lo\`{e}ve basis is typically used when the two matrices in Eq.~(\ref{eq:GE})
are the signal and noise covariance respectively and in that context the eigenvectors
are called signal-to-noise eigenvectors.  In this case, they represent degradation eigenvectors.

The eigenvalues $\lambda_{k}$ are particularly easy to find in our case given the
relationship between the two matrices implies
\begin{equation}
( \mathbf{P}^\mathrm{T} \mathbf{F} \mathbf{P} )(\mathbf{P}^\mathrm{T} \mathbf{F}^{-1} \mathbf{P})\mathbf{v}_{k}= \lambda_{k}  \mathbf{v}_{k},
\end{equation}
or in terms of components
\begin{equation}
    \sum_{\nu}
    \big( \delta^\mathrm{K}_{\mu\nu} - F_{\delta_\br\mu}[\mathbf{F}^{-1}]_{\delta_\br\nu} \big) [{\bf v}_k]_\nu
    = \lambda_k [{\bf v}_k]_\mu.
\end{equation}
For vectors in the
two directions that are orthogonal to $[\mathbf{F}^{-1}]_{\delta_\br\mu}$, the 
second term on the left hand side vanishes and 
 $\lambda_k=1$.   In these directions, marginalization of 
 $\delta_{\br}$ has no impact  since these elements of ${\bf F}^{-1}$
 represent a covariance of the associated direction with $\delta_\br$.  
The remaining eigenvector is in the direction of 
\begin{equation}
[{\bf v}_{{\rm max}}]_\mu \propto F_{\delta_\br\mu}, \vphantom{\Bigg[}
\end{equation}
which is not
in general parallel to $[{\bf F}^{-1}]_{\delta_\br\mu}$ or orthogonal to the 2D space
spanned by the other vectors.  
Thus the direction of maximal degradation 
and the combination of cosmological parameters associated with it 
\begin{equation}
\pi_{\rm max}
= {\bf v}_{\rm max} \cdot \delta {\bf p}
\end{equation}
is not in general the same as those of maximal covariance or degeneracy.
The maximal degradation itself is given by the eigenvalue
\begin{equation}
     \lambda_\textrm{max} = d^2_{\pi_{\rm max}}=  F_{\delta_\br\delta_\br} [\mathbf{F}^{-1}]_{\delta_\br\delta_\br} 
\end{equation}
and is
exactly the degradation in $\sigma_{\delta_\br}$ on marginalizing over cosmological parameters.    
A general linear combination of cosmological parameters
suffers a degradation whose value cannot exceed the maximum since it
is composed partially of the 2  
Karhunen-Lo\`{e}ve 
%Karhunen-Loeve 
directions that are not degraded
 \cite{Smith:2006nk}.

\vfill

\bibliography{sss}

\end{document}